\documentclass[sigconf]{acmart}

\usepackage{graphicx}
\usepackage{subcaption}
\usepackage{tabularx}
\usepackage{longtable}
\usepackage{supertabular}
\AtBeginDocument{%
  \providecommand\BibTeX{{%
    \normalfont B\kern-0.5em{\scshape i\kern-0.25em b}\kern-0.8em\TeX}}}

\copyrightyear{2023}
\acmYear{2023}
\setcopyright{acmlicensed}\acmConference[ASSETS '23]{The 25th International ACM SIGACCESS Conference on Computers and Accessibility}{October 22--25, 2023}{New York, NY, USA}
\acmBooktitle{The 25th International ACM SIGACCESS Conference on Computers and Accessibility (ASSETS '23), October 22--25, 2023, New York, NY, USA}
\acmPrice{15.00}
\acmDOI{10.1145/3597638.3608955}
\acmISBN{979-8-4007-0220-4/23/10}

\newcolumntype{v}{>{\hsize=.8\hsize}X}
\newcolumntype{s}{>{\hsize=.2\hsize}X}

\newcolumntype{t}{>{\hsize=.25\hsize}X}
\newcolumntype{k}{>{\hsize=.75\hsize}X}

\begin{document}

\title[What do Blind and Low-Vision People Really Want from Assistive Smart Devices?]{What do Blind and Low-Vision People Really Want from Assistive Smart Devices? Comparison of the Literature with a Focus Study}

\author{Bhanuka Gamage}
\email{bhanuka.gamage@monash.edu}
\orcid{0000-0003-0502-5883}
\affiliation{%
  \institution{Monash University}
  \city{Melbourne}
  \country{Australia}
}

\author{Thanh-Toan Do}
\email{toan.do@monash.edu}
\orcid{0000-0002-6249-0848}
\affiliation{%
  \institution{Monash University}
  \city{Melbourne}
  \country{Australia}
}

\author{Nicholas Seow Chiang Price}
\email{nicholas.price@monash.edu}
\orcid{0000-0001-9404-7704}
\affiliation{%
  \institution{Monash University}
  \city{Melbourne}
  \country{Australia}
}

\author{Arthur Lowery}
\email{arthur.lowery@monash.edu}
\orcid{}
\affiliation{%
  \institution{Monash University}
  \city{Melbourne}
  \country{Australia}
}

\author{Kim Marriott}
\orcid{0000-0002-9813-0377}
\email{kim.marriott@monash.edu}
\affiliation{%
  \institution{Monash University}
  \city{Melbourne}
  \country{Australia}
}
\renewcommand{\shortauthors}{Bhanuka Gamage, et al.}


\begin{abstract}
  Over the last decade there has been considerable research into how artificial intelligence (AI), specifically computer vision, can assist people who are blind or have low-vision (BLV) to understand their environment.  
  However, there has been almost no research into whether the tasks (object detection, image captioning, text recognition etc.) and devices (smartphones, smart-glasses etc.) investigated by researchers align with the needs and preferences of BLV people. 
  We identified 646 studies published in the last two and a half years that have investigated such assistive AI techniques. 
  We analysed these papers to determine the task, device and participation by BLV individuals. 
  We then interviewed 24 BLV people and asked for their top five AI-based applications and to rank the applications found in the literature. 
  We found only a weak positive correlation between BLV participants' perceived importance of tasks and researchers' focus and that participants prefer conversational agent interface and head-mounted devices.
\end{abstract}





\maketitle

\section{Introduction}

According to the latest World Health Organisation (WHO) statistics, 2.2 billion people have a near or distance vision impairment \cite{whoblindness}.
For sighted people, vision is the primary way they perceive their environment and the objects in it.
Lack of information about their environment and the objects in it is a severe disabling consequence of blindness and low-vision (BLV) \cite{hershberger1992information, butler2002employment}.
As a consequence, a major focus of assistive technology research has been to provide BLV people with information about their environment so that they can travel independently and more confidently interact with the objects and people around them \cite{nearbyexplorer, towardsintelligent, kuriakose2021scenerecog, object_hearing}. 

Recent advances in deep learning have led to breakthroughs in artificial intelligence (AI), in particular in natural language processing (NLP) and computer vision (CV) applications~\cite{vaswani2017attention, chatgpt}.
As a consequence there has been widespread interest in utilising computer vision based assistive technologies to provide BLV people with environmental information.
These range from simple object recognition \cite{bashiri2018object, arora2019real, joshi2020yolo} to autonomous guidance systems \cite{chen2020smart, loomis2005personal, otaegui2013argus} and have been employed on a variety of ``smart'' devices ranging from glasses~\cite{guarese2021cooking, islam2020design, lin2020smart} to robotic dogs~\cite{hong2022development, bruno2019development}.

However, while on average more than 200 papers are now being published each year on this topic, it is unclear if this research actually addresses the real-world needs of BLV people. 
The overarching question we tackle is: \textbf{Does current research in such smart assistive technologies address the needs of the BLV users, and if not what needs should researchers address?}

We conducted a three step investigation to investigate this question.
The first step was a scoping literature review \cite{munn2018systematic} (Section~\ref{section: researcherfocus}). 
We identified 646 research papers published within the past two and a half years that investigated the use of computer vision based applications as an aid for BLV people to comprehend various aspects of their environment. 
We recorded the specific task and devices utilised in these studies, and the kind of interaction model.
We also recorded the involvement of people who are blind or have low-vision (BLV) in the research and their role as prior research suggests that end-user involvement in the design of assistive technology leads to increased satisfaction and adoption by users \cite{steen2011benefits}. 

In the second step of our study (Section \ref{section: communityfocus}), we conducted semi-structured interviews with 24 BLV participants.
During these interviews, we identified the top five tasks that they would like a smart device to assist with, as well as the type of device they preferred. 
We also asked them to rank the usefulness of the tasks identified in the literature. 
Lastly, we inquired about any concerns they had regarding the use of smart devices.

In the final step of our study (Section~\ref{section: analysis}), we compared and contrasted the results of the literature review and the interviews conducted with BLV participants. To the best of our knowledge this is the first research comparing researcher focus with BLV participant priorities in the field of assistive smart devices. Our findings were as follows:
\begin{itemize}
    \item There is only a weak positive correlation between BLV participants' perceived importance of tasks and researchers' focus on those tasks.
    \item The majority of studies (82\%) did not involve BLV participants. However, for those studies that did we found a slightly higher correlation, in particular if BLV participants were involved during the research design stage. 
    \item While all participants were excited by the potential of computer vision based smart assistive technologies, they identified a number of tasks such as filling in paper forms and empty seat detection that warrant more attention by researchers.
    \item Conversational agent type interfaces are preferred by participants, but there is limited research on this topic in the context of smart devices providing environmental information.
    \item Head-mounted devices are favoured by participants, but the choice of wearable device depends on the task and environment.
    \item There is a desire for universal devices or platforms that enable seamless usage without the need for device switching.
\end{itemize}

This study will provide valuable guidance for researchers working in the field of smart assistive technologies and enable them to focus on tasks and interactions that are prioritised by the BLV community. 
In turn this will make it more likely that the technologies developed will be of use and adopted by the community.
\begin{table*}[!t]
    \centering
    \begin{minipage}[t]{0.65\linewidth}
        \centering
        \caption{Keywords}
        \label{tab:keywords}
        \begin{tabular}{|l|l|l|} \hline
            \textbf{Technology} & \textbf{Visual Impairment} & \textbf{Task}   \\ \hline
            "augmented reality" & "sight loss"               & description   \\
            "virtual reality"   & blindness                & caption         \\
            "mixed reality"     & "low vision"               & recognition     \\
            mobile         & "vision loss"              & \textbf{"image enhanc*"}   \\
            wearable*             & "vision aids"              &  \\
             \textbf{"smart*phone app*"} & "impaired vision" &  \\                 
             & "visual loss"   &  \\
                                &  \textbf{"vision impair*"}        &  \\
            \textit{"smart glasses"} & \textbf{"visual* impair*"}            &                 \\ \hline
                                \multicolumn{3}{c}{\textbf{Google Scholar}} \\ \hline
                "smart phone application" & "vision impairment"        &  "image enhancement"\\                                
                "smart phone app" & "vision impairments"       &  "image enhance" \\       
                                &  "vision impaired"     &  "image enhancing" \\   
                                &  "visual impairment"       &   "image enhancer" \\   
                                & "visual impairments"       &                 \\
                                & "visually impaired"       &                 \\ \hline  
        \end{tabular}
    \end{minipage}
    \hfill
    \begin{minipage}[t]{0.33\linewidth}
        \centering
        \caption{Inclusion \& Exclusion criteria}
        \label{tab:criteria}
        \begin{tabularx}{\textwidth}{|X|} \hline
            \textbf{Inclusion}                                                           \\ \hline
            Must utilise computer vision to understand the immediate environment of the user. 
            \\ \hline
            The research application should state it aims to aid the BLV community.              \\ \hline\hline
            \textbf{Exclusion}                                                           \\ \hline
            Devices that are not worn or carried by the user such as smart home devices.
            Applications where only ultra-sonic sensors are used as a mode of detection.     \\ \hline
            Pure VR experiences without any real world integration.                      \\ \hline
            Simulations of different visual impairments using VR.                         \\ \hline
        \end{tabularx}
    \end{minipage}
\end{table*}

\section{Background}
\label{section: background}

In this section we review prior research into the use of smart devices by BLV people.
We define smart devices as personal assistive technologies that use AI-based computer vision to help an individual understand their surroundings.
They might be smartphone applications, stand-alone devices (hand-held, canes) or wearables (belts, caps). We are particularly interested in identifying the tasks that these smart devices can perform to help BLV users.
This can be as simple as detecting a wallet or as complex as guiding the user to buy their groceries.

% Smart devices and ai papers
\subsubsection*{\textbf{Smart device technologies:}} Hundreds of papers have described computer vision based smart devices intended for use by BLV people.
These include hand-held devices \cite{akbari2020vision, lee2021deep, sreeraj2020viziyon}, smart glasses \cite{guarese2021cooking, islam2020design, lin2020smart}, smart caps \cite{konaite2021smart, yang2021lightguide, priya2020machine}, smart canes \cite{yadav2020fusion, akinsiku2020smart, narayani2021design} and many more \cite{kathiravan2021smart, tachiquin2021wearable, arguello2022belt, boldini2020piezoelectric}.
These utilise image enhancement \cite{zhao2016cuesee, lang2020augmented, du2021automated}, question and answering \cite{mobilevqa, yamanaka2022one} and other forms of interaction  to assist their users.
Tasks include Face Detection \cite{facerecog2, facerecog1}, Text Recognition \cite{textrecog1, textrecog2}, Road Sign Detection \cite{roadsign1, roadsign2} and many more.
Numerous commercially available AI smart devices have also emerged employing a variety of devices~\cite{seeingai, orcam, envisionglasses, straptech, nueyes, visionbuddy, googlelookout}.

% Review papers
Several papers have reviewed these different devices and technologies \cite{li2022scoping, patel2020assistive, simoes2020review, kuriakose2020multimodal, khan2021insight, valipoor2022recent}. 
Li et al. \cite{li2022scoping} evaluated head-mounted displays (HMD) for assistive and therapeutic applications for BLV people.
They reviewed 61 studies and classified the studies based on the type of HMDs, the visualisation approaches, the visual condition and if the evaluation was done using a user study.
The main focus of their study was to analyse the current state of assistive HMDs and present emerging trends.

Patel and Parmar \cite{patel2020assistive} reviewed 13 vision substitution devices and studies for BLV individuals that used computer vision and image enhancement.
The authors noted that only a limited number of assistive devices were commercially available, and the majority of devices had not been tested under real-world conditions.

Khan and Khusro \cite{khan2021insight} conducted a review of smartphone-based assistive technologies for BLV people, emphasising the importance of technological advancements, inclusive interfaces, and collaborative efforts among medical specialists, computer professionals, usability experts, and domain users to maximise the potential of ICT-based interventions for the visually impaired.

Sim{\~o}es et al. \cite{simoes2020review} reviewed technologies and methods for assisting BLV people with indoor navigation.
They classified these indoor positioning systems and discussed the advantages and disadvantages of each approach.

The review of Kuriakose et al. \cite{kuriakose2020multimodal} focused on studies into multi-modal navigation for BLV people.
They provide a comprehensive review of the different forms of devices and their modalities.
In each of the studies, they also identified if the system was evaluated with BLV users.

In their systematic mapping review of 105 papers, Valipoor and Antonio \cite{valipoor2022recent} primarily centred their investigation on the object recognition, obstacle detection and depth detection aspects of assistive solutions. 
The review highlighted that recent advances in computer vision hold the potential to revolutionise the creation of advanced and high-quality assistive technologies for BLV users.

However, none of these surveys directly considered the needs of the BLV community. 

% User needs with reviews in other areas.
\subsubsection*{\textbf{User requirements:}} Only a small number of papers have explicitly examined  the requirements of BLV users.
In 2001, Duckett and Pratt \cite{participant_importance} conducted a study to identify the higher-level research topics deemed important by BLV participants. Understanding the environment was one of the main topics identified. 
%to investigate the opinions of the BLV community regarding assistive technology research. 
% In 2001, Duckett and Pratt \cite{participant_importance} conducted a study to investigate the opinions of the visually impaired community regarding research on visual impairment.
%The authors focused on identifying higher-level research topics deemed important by participants, and found that access to the environment was one of the main topics identified. 
However, this study was conducted before the widespread adoption of smartphones and mobile devices and did not specifically consider AI-based smart devices.

Tapu et al. \cite{tapu2020wearable} investigated the perception of wearable assistive devices by the BLV community. 
They reviewed and categorised these devices into sensor based and camera based devices.
The authors then conducted consultations with visually impaired individuals, researchers, and software developers to identify a set of features that a wearable assistive device must possess to gain acceptance within the BLV community.
Processing speed, portability, robustness and friendliness were some of them.
They then scored the devices in the review with these features and found that none of them satisfied all the criteria.

In 2020, Plikynas et al. \cite{plikynas2020indoor} explored requirements for indoor navigation systems.
They looked into 27 papers with different forms of hardware and sensors and proposed directions for future developments.
During their interview stage, they asked blind participants about the five biggest problems when orientating/navigating indoors.
Finding room numbers, finding elevators and reading numbers at the bank were the top three problems.

The most relevant study is that of Golubova et al. \cite{golubova_design_2021}. 
They identified tasks that BLV participants would like a smart sight aid to perform.
They did by asking the participants to wear recording glasses that can be turned on to capture a moment where they would use a `perfect sight aid`.
Reading package labels, reading signs, reading print on TV and identifying medicine were some of the most reported situations.
One insight from the study was that many participants prefer devices that can perform a spontaneous task over a device that they have to wear for long hours.
A limitation of the study was that it implicitly focused on HMD devices and did not explore participant preferences for smart devices or ways of interacting with these devices. 

% the gap
To date we see that there has been no comparison between the tasks and smart devices that researchers have focused on and the needs of BLV people. 
Furthermore, apart from~\cite{golubova_design_2021} there has not been a general investigation into the tasks that BLV people would like smart devices to perform, and no investigation into the type of smart devices they prefer and the kind of interaction they would like. 
These are the issues that we address in this paper.
\section{Study 1 - Scoping Review}
\label{section: researcherfocus}

The first step in our investigation was a literature review to ascertain the various tasks and smart devices that researchers have focused on. 
As part of this review we also examined the extent to which BLV participants were involved in the research. 
Given our goal of exploring and mapping the broader topic of smart assistive devices based on technology, devices, interactions, and involvement of BLV individuals, we opted to conduct a scoping review. 
As outlined by Munn et al. \cite{munn2018systematic}, scoping reviews are particularly well-suited for examining how research is conducted on a specific topic and identifying and analysing gaps in the knowledge base, and identifying the types of available evidence in a given field.

\subsection{Literature Search and Review Methodology}

We captured papers from four databases: Google Scholar, Scopus, Web of Science, and PubMed. 
Search was based on keywords capturing three criteria: Technology, Visual Impairment, and high level Task. 
The top table of Table~\ref{tab:keywords} shows the keywords: keywords in the same column were combined with "OR" and the columns combined with an "AND" operator during the search process.

Our initial search was conducted in July 2022. We restricted our search to the past two and a half years, specifically from 2020 to July 15th, 2022. 
There were two reasons for this: First, the majority of AI-based smart device research has been carried out in recent years. Second,the sheer size of the literature--we initially looked at the last five years and found over 30,000 papers.

\subsubsection*{\textbf{Initial search:}}
Our initial search with the keywords from the upper table of Table~\ref{tab:keywords} with Google Scholar, found 4,992 papers, and that with Scopus, Web of Science, and PubMed, resulted in 264 papers.
After applying a title and abstract screening process based on the exclusion criteria outlined in Table \ref{tab:criteria}, we narrowed down the results to 454 papers from Google Scholar and 194 papers from the other databases. Merging and removal of duplicates, resulted in 587 papers for full-text review. 
During this review process, we excluded additional papers that  did not meet our exclusion criteria, resulting in a final set of 477 papers.
Selection of papers were primarily carried out by a single researcher but in consultation with other members of the research team when unclear.

\subsubsection*{\textbf{Followup search:}} After conducting the initial search we discovered a limitation in Google Scholar's ability to handle wildcard queries within quoted multi-word queries.
Unlike other databases, Google Scholar appears to interprete the symbol `*' as a wildcard word within multi-word queries, rather than as a wildcard completion. 
This means that, for example, that a search for "image enhanc*" will match "image enhance by" but only match "image enhancing" inconsistently on Google Scholar, while this term will consistently match "image enhancement" on the other databases.

As a solution, we expanded each multi-word-wildcard query (bolded text in Table \ref{tab:keywords}) into separate multi-word keywords, as shown in the Google Scholar section of Table \ref{tab:keywords} and repeated the Google Scholar search.

We identified 11,034 new papers that were not in the initial search.
Since the second search was conducted after July 15, 2022, we also removed papers that were published after that date to ensure consistency between the two groups. 
Filtering by the date and applying the title and abstract screening resulted in 214 papers.
After merging the new papers with the previous results, we found an additional 140 papers, bringing the total number of extracted papers from 477 to 617.

Furthermore, it came to our attention that the term \textit{"smart glasses"} was inadvertently omitted as a specific keyword. To rectify this, a third search was conducted, resulting in the inclusion of 29 additional papers, bringing the total count from 617 to 646.

\subsection{Data Extraction} 
\label{sec:data-extraction}

The following information was recorded for each paper.
\begin{itemize}
    \item \textbf{Task:} The low-level task(s) the research was designed to support.
    \item \textbf{Device:} Type of smart device used in the research.
    \item \textbf{Interaction model:} The high-level interaction model(s) from the user's perspective. We examined whether the device enhanced images for the user (Image Enhancement), identified and described the user's surroundings (Scene and Object Description), found or located objects requested by the user (Detection and Recognition of User-Requested Objects), or engaged in conversation with the user (Question Answering).
    \item \textbf{Involvement of BLV participants:} We determined whether BLV participants were involved or not. And, if they were, whether this was during the design phase, the evaluation phase, or both phases.
\end{itemize}

\subsection{Results}

\subsubsection*{\textbf{Tasks}}

Low-level tasks were extracted from the research studies. 
Two researchers went through these and combined tasks if they were closely related. 
For example, a study was detecting fruits and vegetables and another detecting seafood, were combined into Food Recognition.
It should be noted that some tasks encompass other more specific tasks. For instance, Drug Pill Detection and Indoor Sign Detection utilise Text Recognition, but were classified as distinct tasks because they involved additional classification steps to provide more comprehensive output to the users, rather than simply reading text.

After finishing the data extraction, two of the researchers categorised these low-level tasks into higher-level groups.
The initial grouping was informed by the Australian Assistive Products for Persons with Disability Classification and Terminology \cite{australia_standard}, which is based on the WHO ISO Assistive Technology Standards and Guidelines \cite{iso_2022}. 
We introduced a new category called "Assistive products for cultural and sports activities" as we felt that tasks in this category did not readily fit into the other categories.
The low-level tasks, their description and high-level groups are shown in Table \ref{tab:tasks} (in the Appendix). 

\begin{figure*}
    \centering
    \includegraphics[width=0.82\textwidth,keepaspectratio]{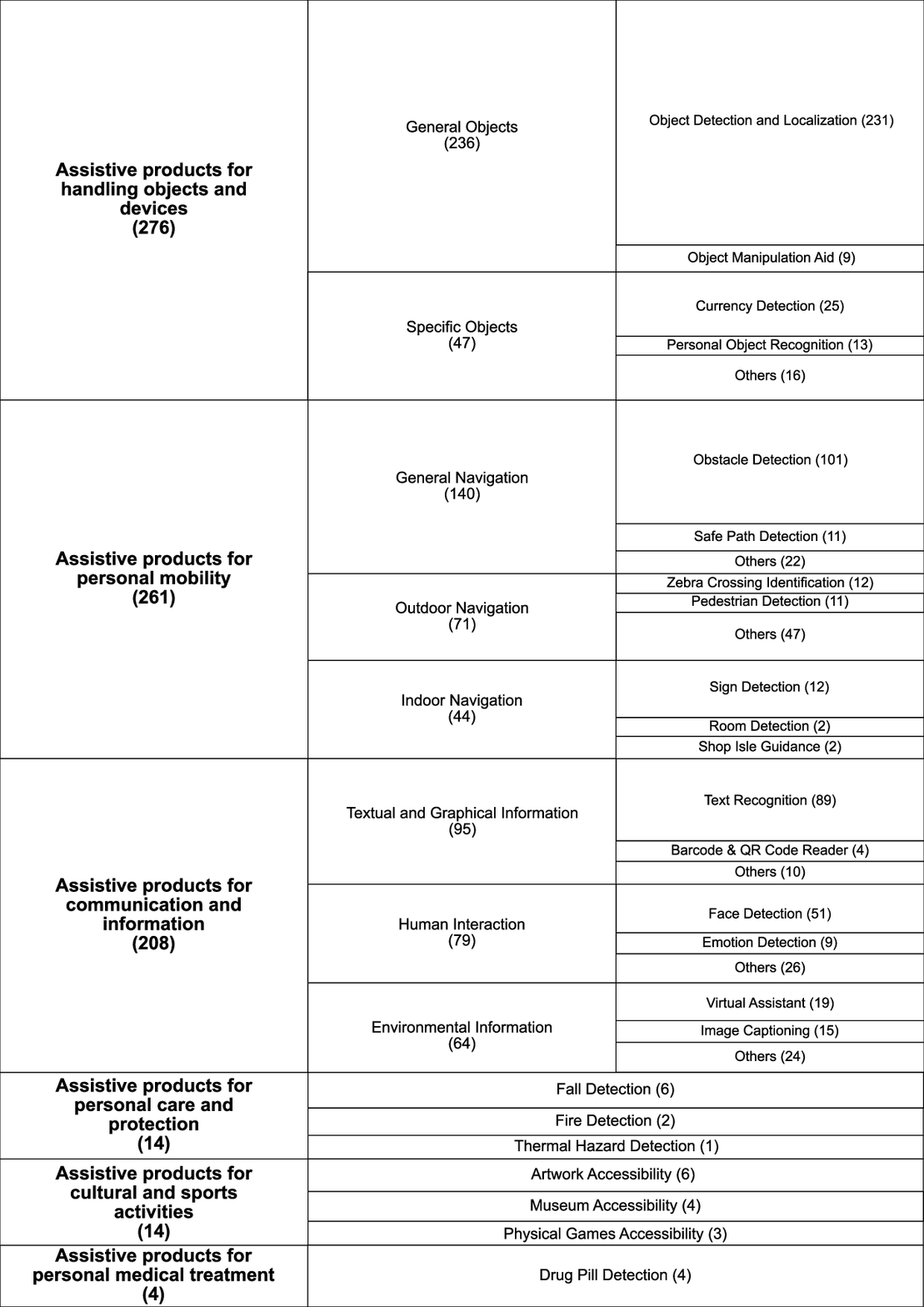}
    \caption{Breakdown of the counts of papers identified for each types of tasks. Note the height of each sub-box is only indicative of the number of papers in that category.}
    \label{fig:task-counts}
    \Description{This figure is an Icicle Plot that shows the counts of papers that were identified for each task group in the review, categorised by the high-level groups. The trends are described in the 4th paragraph Section 3.3 Results. Refer to Figure1.csv in supplementary material for all the data.}
\end{figure*}

Once the tasks were determined we counted the number of papers addressing each task. 
The counts are summarised in Figure \ref{fig:task-counts}. 
It is worth noting that some papers addressed multiple tasks, such as Currency Recognition and Face Detection. 
Therefore, the total number of papers for all high-level categories adds up to more than 646 papers.

Our count revealed that the category of assistive products for handling objects and devices had the highest number of papers, accounting for 42.7\% of the studies, followed by assistive products for personal mobility (40.4\%), and communication and information (32.2\%). 
The Object Detection \& Localization task had the highest number of papers. 
The majority of these studies we reviewed were trained on well-known datasets such as COCO \cite{lin2014microsoft} and ImageNet \cite{deng2009imagenet}, which include hundreds of object classes.
As a result, these devices detect and share all objects from their training dataset, regardless of their relevance to the user.

\subsubsection*{\textbf{Device}}

We collated a list of smart devices used in the studies and counted the number of papers that used each device.
In cases where smart devices ran on Raspberry Pi or similar computing devices but did not specify the wearable location, we categorised it as a body-mounted device. 
Additionally, we classified smart-glasses and devices with head-mounted cameras as head-mounted devices.
It is worth noting that some studies may fall under more than one category, such as a head-mounted device that utilises a smartphone app for voice input. 
In such cases, the study is counted for both wearables. 

\begin{figure*}
        \centering
        \includegraphics[width=15cm,keepaspectratio]{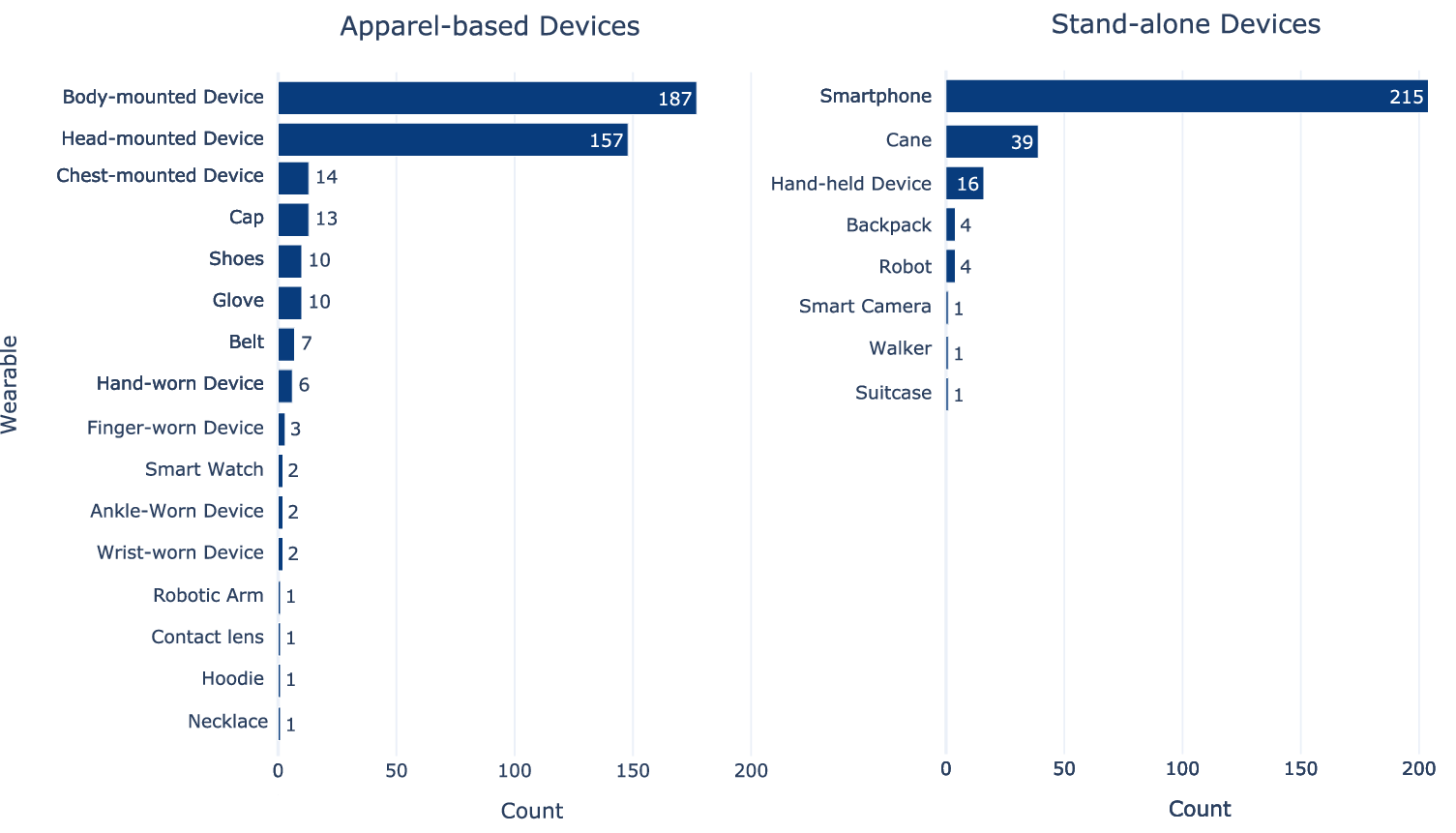}
        \caption{Count of papers identified for the types of devices}
        \Description{This figure shows two horizontal bar charts next to each other with titles: Apparel-based Devices and Stand-alone Devices. It shows the count of papers in the review for each type of wearable device.For apparel-based devices, the highest to lowest are: Body-mounted Device 187,Head-mounted Device 157,Chest-mounted Device 14,Cap 13,Shoes 10,Glove 10,Belt 7,Hand-worn Device 6,Finger-worn Device 3,Smart Watch 2,Ankle-Worn Device 2,Wrist-worn Device 2, Robotic Arm 1,Contact lens 1, Hoodie 1, Necklace 1. For stand-alone devices, the highest to lowest are: Smartphone 215,Cane 39,Hand-held Device 16,Backpack 4,Robot 4,Smart Camera 1,Walker 1,Suitcase 1}
        \label{fig:wearable-model-count}
\end{figure*}

Figure \ref{fig:wearable-model-count} depicts the number of studies conducted for each type of wearable device. 
We observed 24 distinct types of devices in our review, with only 9 of them having at least 10 studies.
The majority of studies employed smartphones as the primary device. This can be attributed to the fact that smartphones have all three essential components (i.e., camera, microphone, and computing unit) integrated into a single device as identified by Plikynas et al. \cite{plikynas2020indoor} as well. 
Additionally, the widespread adoption of smartphone software development kits (SDKs) may have played a role in their popularity compared to more specialised tools required for other wearables.

\begin{figure*}
        \centering
        \includegraphics[width=15cm,keepaspectratio]{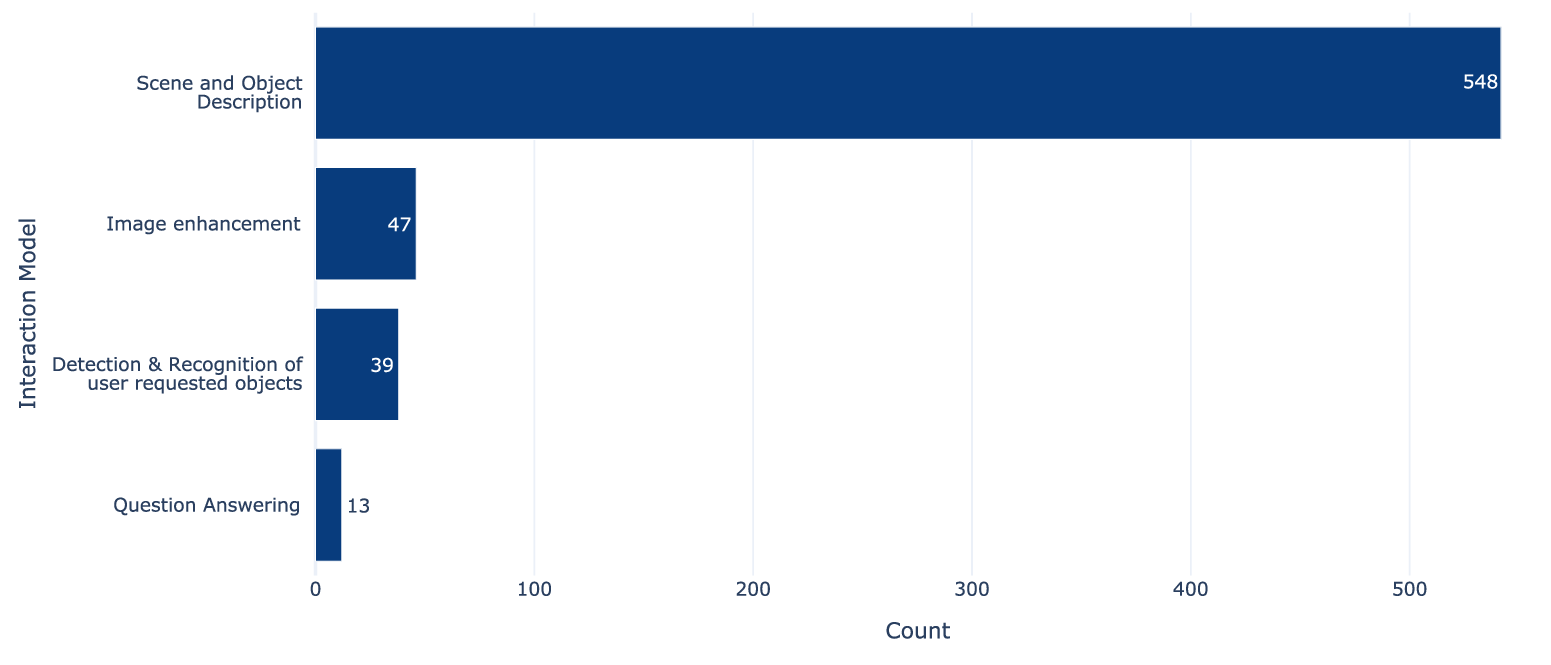}
        \caption{Count of papers identified for each interaction model}
        \Description{This figure shows a horizontal bar chart with the count of papers identified for each of the interaction models. The interaction models are sorted from the highest to lowest as follows: Scene and Object Description 548,Image enhancement 47, Detection \& Recognition of user requested objects 39,Question Answering 13}
        \label{fig:interaction-model-count}
\end{figure*}

\subsubsection*{\textbf{Interaction model}}
The majority of studies (85\%) focused on Scene and Object Description, while Question and Answering received the least attention (as shown in Figure \ref{fig:interaction-model-count}).
This could be attributed to two reasons. 
Firstly, wearable devices have limited computing power, which poses a challenge for developing on-device Question and Answering systems. 
Secondly, many studies were focused on the quantitative performance of the computer vision algorithms rather than user interaction.

\subsubsection*{\textbf{Involvement of BLV participants}}

\begin{figure*}
    \centering
    \includegraphics[width=15cm,keepaspectratio]{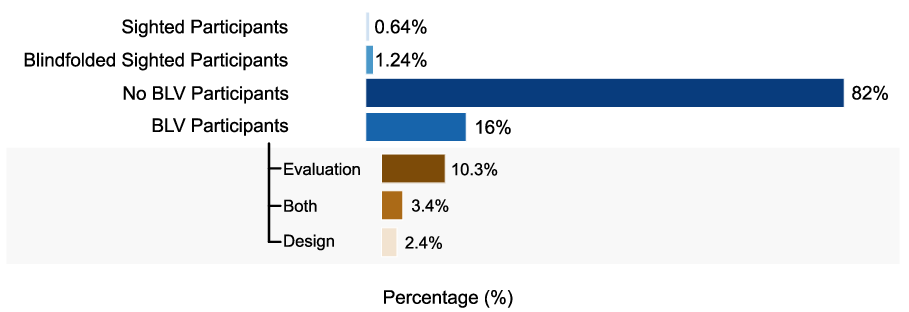}
    \caption{Type of participant involvement in research studies and the stages of involvement}
    \Description{This figure shows two horizontal bar charts with the second one extending from the last row (‘BLV participants”) in the first chart. The types of participants are listed as follows: Sighted Participants 0.64\%,Blindfolded Sighted Participants 1.24\%,No BLV Participants 82\%,BLV Participants 16\%.For the studies with BLV participants, the values are sorted from highest to lowest as follows: Evaluation 10.3\%, Both 3.4\%, Design 2.4\%}
    \label{fig:blind-involvement}
\end{figure*}

After collating the results, our analysis of the reviewed papers revealed that the BLV participants were not involved in the majority (82\%) of the studies and only 105 out of the 646 reviewed had BLV participants--see Figure \ref{fig:blind-involvement}.
We further examined the involvement of BLV participants at different stages of the studies.
We found that in the majority of papers, BLV involvement was restricted to evaluation and only 38 papers out of the 646 reviewed involved BLV participants in the design, ideation, or requirement gathering stage. 

\section{Study 2 - Needs of People Who are Blind or have Low-Vision (BLV)}
\label{section: communityfocus}

The second step in our research was to conduct semi-structured interviews with BLV participants to ascertain what they thought were the most useful tasks that smart devices could be used for. 
As part of this we asked about preferred interaction models and devices as well as asking them to rank the importance of the tasks that had been studied by researchers. 

\subsection{Participants}

We conducted semi-structured interviews with 24 participants who either have low-vision or are blind. 
These lasted approximately one hour and participants were compensated with a \$50AUD gift card for their time. 
Of the 24 participants, 10 had low-vision and 14 were blind, with varying characteristics across age groups and age of onset.
Table \ref{tab:participant_details} (in the Appendix) summarises the demographic details of the participants, including age, level of vision, age of onset, and level of tech adoption. 

\subsection{Procedure}

The interview had 5 parts.

\subsubsection*{\textbf{Demographic information:}}
At the beginning of each interview, a series of general questions were asked to ascertain the participants' demographic information and visual condition. 
Additionally, participants were requested to rank their level of technology adoption, which was categorised according to the summary provided by \cite{beal1957diffusion} as shown in Table \ref{tab:technology-adoption} (in the Appendix).

\subsubsection*{\textbf{Current technology use:}}
Next, we asked about the types of devices participants currently utilise in their daily routines. 
If a participant mentioned using a smartphone or tablet, we further inquired about the applications they used on these devices. 

Then we asked participants whether they have utilised any services that provide sighted assistance through video conferencing, such as AIRA \cite{aira} and Be My Eyes \cite{bemyeyes}. 
For those who have used such services, we followed up with questions to identify the tasks for which they utilised these services. 
Our objective was to identify tasks for which BLV individuals currently require sighted assistance, with the goal of determining if smart devices could potentially assist with these tasks.
Lastly, we inquired whether any of the participants had any concerns regarding these services.

\subsubsection*{\textbf{Most useful tasks:}}Prior to asking participants to list the tasks they felt would be the most useful, we provided them with a brief description of the types of smart devices, including smartphones, head-mounted devices, and chest-mounted devices. 
Additionally, we explained how artificial intelligence and machine learning has advanced in terms of mimicking human intelligence and how it can be incorporated into these devices. 
This was done to ensure that the participants had a general understanding of what constitutes a smart device. 
We carefully refrained from sharing any details that we identified from the review about the different tasks that these devices could perform, as we first sought to obtain the participants' perspective before presenting our findings from the review.

Next, we asked participants to list their top five tasks that they would like to accomplish using a smart device, followed by their suggestion for the type of smart device that would be best suited to the task. 

In order to understand the two forms of modalities, i.e., human-computer (input) and computer-human (output) \cite{mittal2011versatile}, we asked participants how they would prefer to interact with the device, including input methods such as voice, tactile buttons, and touch screen, and how they would prefer the device to provide information to them through output methods such as audio, visual or vibrations.

\subsubsection*{\textbf{Ranking of researcher tasks:}} 
After this, participants were asked to rank the various tasks that researchers had focused on, as described in Table \ref{tab:tasks}. 
To keep the interview to around one hour we could not ask each participant about all of the tasks. Instead we divided the participants into four groups, with each group ranking 18 tasks. 
We selectively assigned participants to each group to balance out the representation of low-vision and blind participants, ensuring that each group had at least two low-vision participants. 
To prevent any potential biases that may arise from the order in which the 18 tasks were presented to the participants during the interview, we randomised the task order for each individual.

While reviewing each task, the participants were asked to rank its importance based on its usefulness, ranging from "not useful", "slightly useful", "very useful", to "invaluable." 
Following this, they were asked a few questions to understand why the task was significant as the reason might differ from the researchers' original intention. 
Finally, they were requested to choose the `best' type of smart device that would fulfil the task.

After the completion of the task ranking process, we provided the participants with an opportunity to re-evaluate their initial top five tasks to ensure that they had the chance to adjust their priorities based on the insights they gained from the ranking process.

\subsubsection*{\textbf{Overall perception of smart devices:}}
Following this, we asked the participants to rank the overall usefulness of smart assistive devices on the same scale as before, in order to gain a better understanding of their general perception of these devices. 
Lastly, we followed up with a question to discuss any concerns they may have regarding the development of smart devices for BLV individuals.

\subsection{Results}

\subsubsection*{\textbf{Current technology use:}} \label{sec:current-technology-use} All participants currently use a smart phone.
This is consistent with Plikynas et al. \cite{plikynas2020indoor}, where they identified smartphones to be the most common electronic travelling aid.
OrCam (3), Envision glasses (2), BuzzClip (1), StrapTech (1), MiniGuide (1) and Trekker Breeze (1) were some of the other devices used. Note that the number in () is the number of participants using the device.
We also found that Seeing AI (14), Google/Apple Maps (13), Soundscape (8), Be My Eyes (7) and Envision (6) were the most commonly used smartphone applications.

Out of the 16 participants who have used video conferencing services for sighted assistance, we identified the most frequently requested tasks. 
Reading text was the most commonly requested task by both blind and low-vision participants with 10 participants requesting it. 
Other frequently requested tasks included colour detection, website accessibility, navigating new environments, cleaning broken glass, framing and taking photographs, finding personal items, and assessing their physical presentation.

The results also revealed that our blind participants use these services more than low-vision participants (11 out of 14 blind participants use these services compared to 5 out of 10 LV participants).
Out of the participants who used these services the blind participants tended to favour using AIRA over Be My Eyes, while low-vision participants preferred Be My Eyes over AIRA. 
This could be explained by the fact that low-vision individuals may rely on their limited vision to assist them with tasks, and the crowd-sourced support workers used by Be My Eyes may be sufficient for their needs. 
In contrast, blind individuals may require more specialised assistance and may therefore prefer the professionally trained staff provided by AIRA.

Finally, the majority of participants expressed positive feedback regarding the sighted assistance services, appreciating their convenience and professionalism, particularly when they lacked physical assistance.
Only two participants raised concerns regarding issues such as privacy, connectivity, and high costs.

\subsubsection*{\textbf{Most useful tasks:}} \label{sec:important_to_participants}
We compiled the responses of participants regarding the top 3-5 most preferred tasks and mapped them to the lowest level task classification shown in Table \ref{tab:tasks}. 
For instance, one participant mentioned "something that can identify people" which was then mapped to Face Detection.

\begin{figure*}
    \centering
    \includegraphics[width=15cm,keepaspectratio]{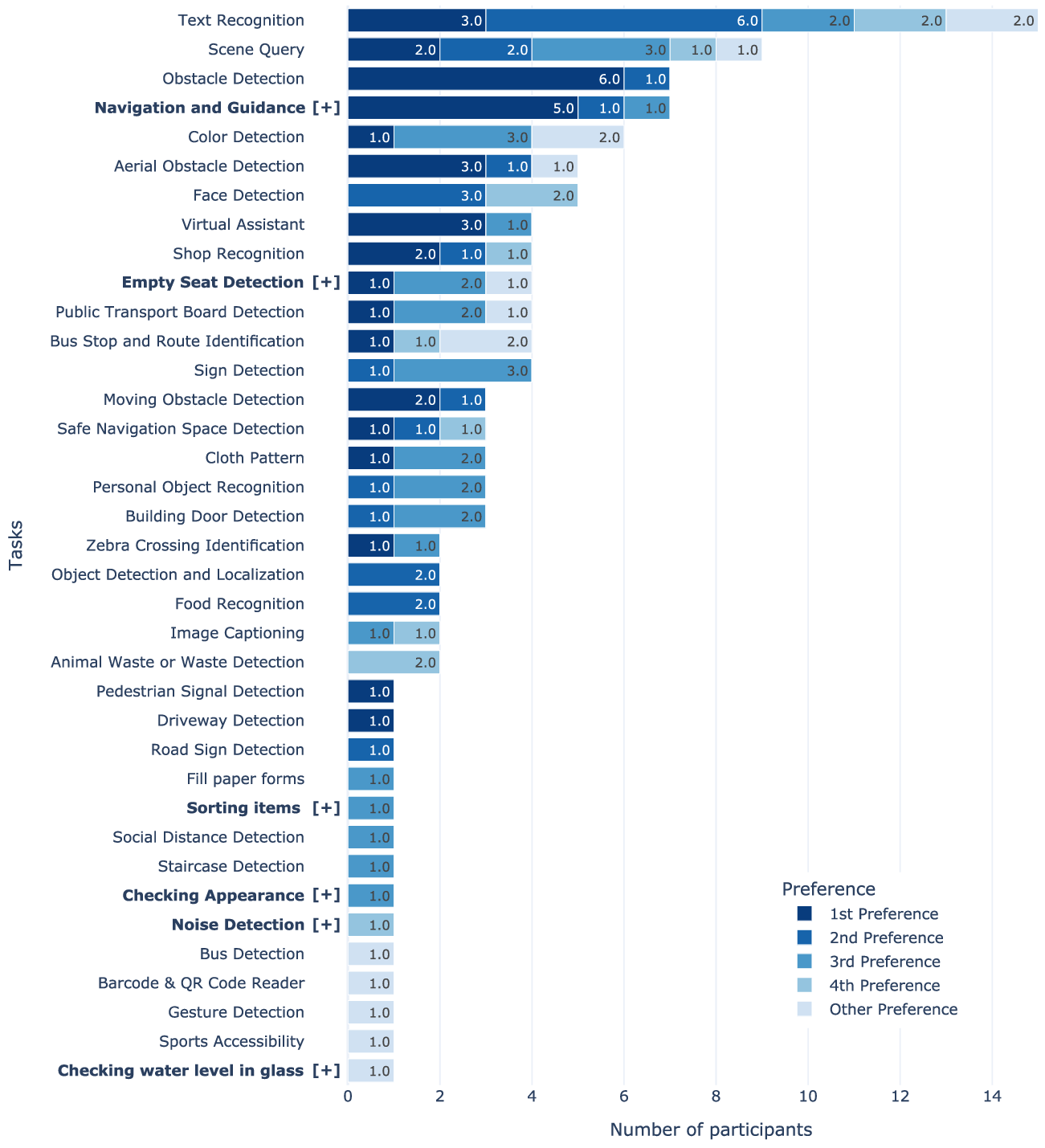}
    \caption{Participants' Five Most Useful Tasks Ranked by Order of Preference; The identified tasks are presented in order from most to least preferred, with tasks not listed in our review task list indicated in bold font with a [+] symbol next to them.}
    \Description{This figure shows a stacked horizontal bar chart, with each segment showcasing the preference type. The trends of the plots are described in the second paragraph of the Most useful task subsection in Section 4.3 Results. Refer to Figure5.csv in supplementary material for all the data.}
    \label{fig:user-preference}
\end{figure*}

The most preferred task was Text Recognition, which was indicated by 15 out of the 20 participants, as demonstrated in Figure \ref{fig:user-preference}.
Many participants reported that when they are outdoors, they often have to rely on others to read text for them, indicating the need for a reliable smart device to perform this task.
The results of our study align with the findings of Golubova et al. \cite{golubova_design_2021}, who reported that 18 out of their 32 participants identified reading any type of text as the most commonly named activity of importance.
This finding also aligns with the results described in Section \ref{sec:current-technology-use}, where reading text was found to be the most requested task from video conferencing services.

We also identified that Obstacle Detection had the highest number of first preferences, with six participants selecting it. 
One participant, P9, reported that Obstacle Detection while navigating was the most challenging and stressful task for them. 
Therefore, a smart device that could aid in this area would greatly enhance their independence.

Figure \ref{fig:user-preference} highlights several tasks in bold, including Empty Seat Detection, Sorting Items and Checking Appearance, that were not included in the list of tasks from the review. 
We also included Navigation and Guidance as a separate task though it does overlap with Safe Path Navigation. 
This was primarily due to participants mentioning the need for Night Time Navigation and Guidance, which was not fully captured under Safe Path Detection. 
It is also because Navigation and Guidance is a higher-level task that combines several low-level tasks, such as Shop Recognition, Building Door Detection, and Safe Path Detection, to provide seamless interaction with the user, as demonstrated by \cite{suitcase_roller}. 

We could not identify differences in the choice of top five tasks between the blind and low-vision participants.

\subsubsection*{Participant's preferred interaction mode:}

\begin{figure*}
    \centering
    \includegraphics[width=15cm,keepaspectratio]{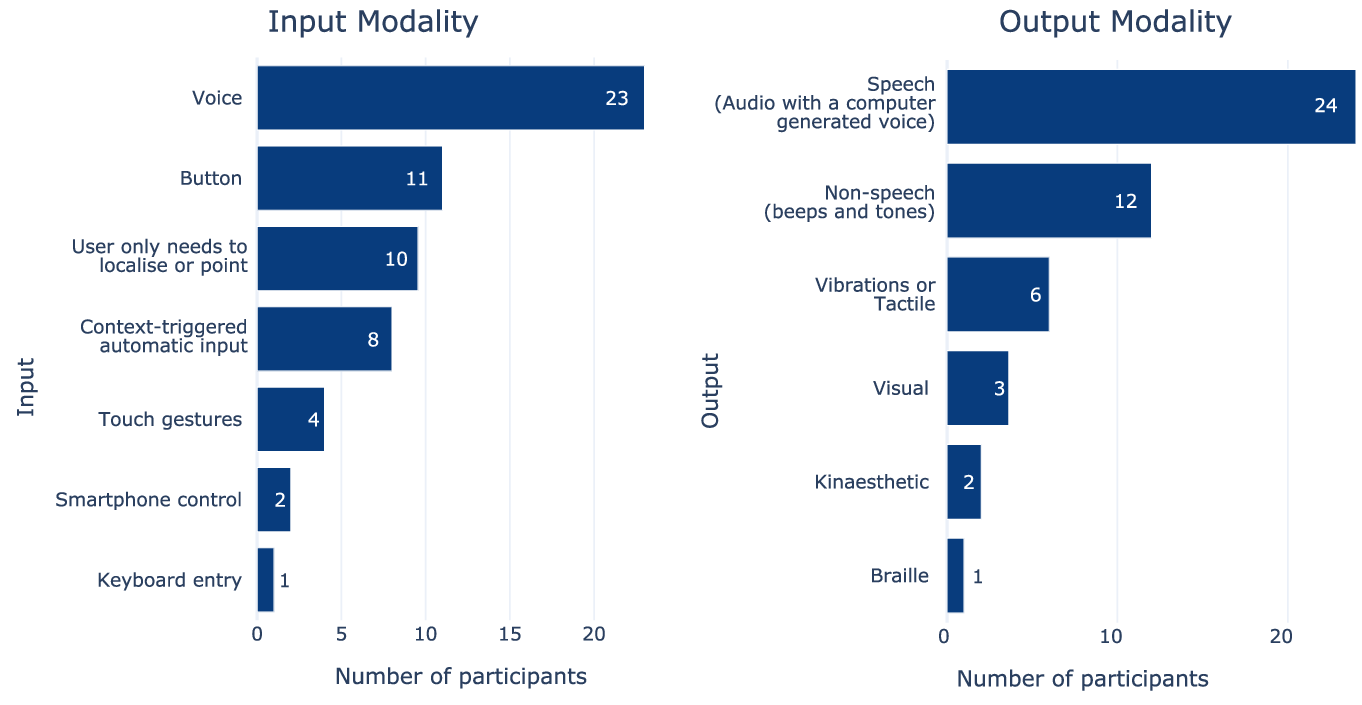}
    \caption{Interaction modes identified in the interview with the number of participants that preferred it for at least one of their top five tasks}
    \Description{This figure shows two horizontal bar charts next to each other with titles: Input Modality and Output Modality. For Input Modality, the highest to lowest are: Voice 23,Button 11,User only needs to localise or point 10,Context-triggered automatic input 8,Touch gestures 4, Smartphone control 2,Keyboard entry 1. For Output Modality, the highest to lowest are: Speech (Audio with a computer generated voice) 24,Non-speech (beeps and tones) 12,Vibrations or Tactile 6, Visual 3,Kinaesthetic 2,Braile 1}
    \label{fig:input-output-count}
\end{figure*}

Figure \ref{fig:input-output-count} shows the different input and output modalities identified by the participants. It gives the number of participants that stated this modality for at least one of their tasks.
In terms of preferred modes of interaction, 23 out of the 24 participants expressed a preference for voice input for at least one of their top tasks. 
This finding is also reflected in \cite{speechreview} where they compare the use of voice input by blind participants over sighted participants and found that blind individuals use voice input more frequently.
\cite{speech_input_blind} also found that 6 out of 8 blind participants preferred voice input over on-screen keyboard on mobile devices.

Participants also noted that, in the case of a smart glass device (head-mounted), voice input would be significantly more convenient for issuing queries, whereas a handheld device might be better suited to button-based input.
Other popular input modalities included, `user only needs to localise' and `context-triggered automatic input' and button input.
P4 described their ideal face-detection smart glass as having the capability to automatically track individuals they interacted with, capturing their names and periodically reminding the user when a known or unknown person was addressing them.

All participants indicated a preference for speech output as the primary means of interaction for at least one of their top tasks.
Four participants expressed a preference for non-speech modalities such as beeps and tones, as well as vibration, due to concerns that speech output might interfere with their reliance on surrounding environment sounds.
This finding aligns with the findings of \cite{plikynas2020indoor} who found that participants did not want a device that would block their hearing when navigating indoor. However a factor impacting on the use of speech output is the need to avoid overwhelming the user.
In particular, one participant (P14) emphasised the importance of the device being clear and concise when communicating with the user, as an overly verbose device could be overwhelming and interfere with task performance.

Comparing between the blind and low-vision groups, for input modalities, 7 blind participants showed a preference for the `user only needing to localise`, while only 3 low-vision participants exhibited this preference. 
In terms of output modalities, 3 low-vision participants preferred the visual modality, whereas (as one might expect) none of the blind participants did so.

\subsubsection*{Participant's preferred device:} \label{met-results:preferred-device}
As depicted in column B in Figure \ref{fig:wearable-types}, the top three wearable types preferred by the participants are head-mounted devices, smartphones, and hand-held devices.
It is important to note that during the study, participants utilised the terms "hand-held devices" and "smartphones" interchangeably, where as certain participants explicitly referred smartphone applications.
The participants highlighted numerous advantages of using smart glasses (head-mounted devices) over smartphones and hand-held devices.
A key advantage of smart glasses is that they follow the natural gaze of the user's eyes and head, which eliminates the need for awkward pointing or manoeuvring to capture what is in front of them, as noted by P22:
\begin{quote}
\textit{"Generally people are looking at what they are doing, so it's natural to hold something up to the face, to look at it and feel it, so there by having it there [head-mounted], having it hands free, having not to worry about holding something."}
\end{quote}

Smart glasses also offer an additional benefit over hand-held devices in that they can be used without requiring the use of both hands. 
This is especially advantageous for guide dog users who may already be using one hand to hold the guide dog. 
P1 noted that: 
\begin{quote}
\textit{I have a guide dog that I'm out and about with, so that means that one of my hands is always busy with the dog.}
\end{quote}

Finally, our comparison between the blind and low-vision group found no discernible preference in the type of device for their top five tasks.

\begin{figure*}
    \centering
    \includegraphics[width=15cm,keepaspectratio]{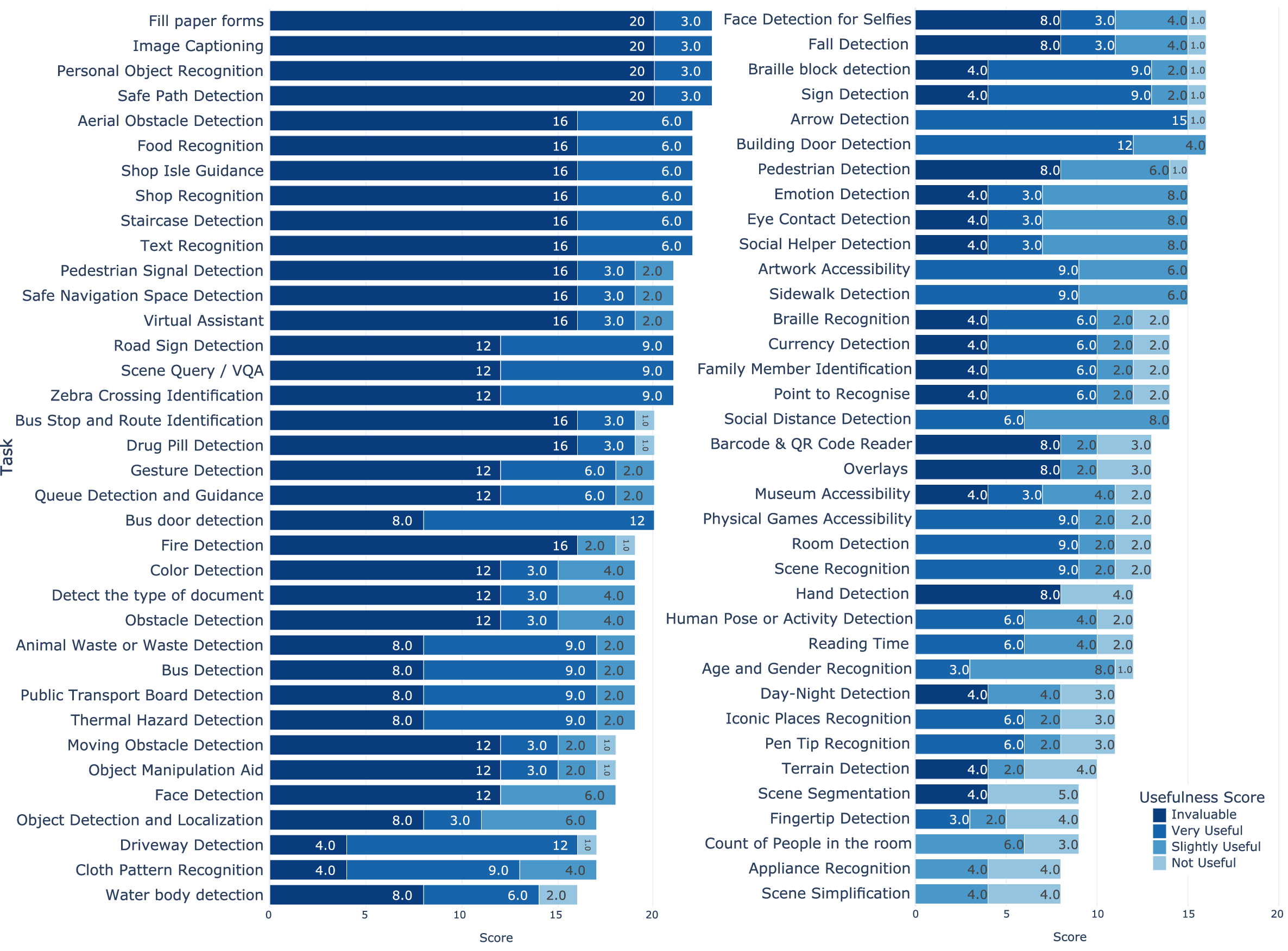}
    \caption{Ranking of tasks identified in the review based on participants' score of usefulness, ranging from "Not Useful," "Slightly Useful," "Very Useful," to "Invaluable". Each category was assigned a score from 1 to 4 and the tasks are then sorted based on the total score. The figure is presented from left to right and top to bottom.}
    \Description{This figure shows a stacked horizontal bar chart, with each segment showcasing the usefulness score shared by the participants. The trends of the plots are described in the first paragraph of the Ranking of researcher tasks subsection in Section 4.3 Results. Refer to Figure7.csv in supplementary material for all the data.}
    \label{fig:usefulness-score-rank}
\end{figure*}

\subsubsection*{\textbf{Ranking of researcher tasks:}}
Figure \ref{fig:usefulness-score-rank} shows the ranking assigned by participants to the tasks identified in the survey. 
Image Captioning, Personal Object Recognition, Safe Path Detection and Fill Paper Forms all received the same highest score. 
Several participants also expressed the need for a smart device that could assist them in filling out paper forms, as they currently have no alternative but to rely on a sighted person.
One participant, P4, emphasised the importance of accurate Image Captioning, as it renders them unable to comprehend the content without it.

Appliance Recognition and Scene Simplification were ranked as the lowest in terms of usefulness.
One reason why Appliance Recognition was not useful, as pointed out by P4, was that she sets up her living space in a way that allows her to easily identify appliances, making a smart device redundant in this regard. 
P3 suggested that Scene Simplification may not be useful as he did not want to rely on a device that uses vision, given the knowledge that his vision would deteriorate over time.

In the course of our interview, we made an intriguing observation regarding the tasks extracted from the paper \cite{abirami2020customized}, which aimed at developing a system for detecting Animal Waste or Waste to avoid stepping on it. 
Upon being asked during the interviews, two out of six participants stated that their primary use for the device would be to locate the animal waste left by their guide dog. 
This is a significant consideration, as visually impaired individuals face challenges in picking up after their guide dogs. 
They explained that if the dog defecates outside, they may have no choice but to leave the waste as they are unable to locate it. 
This suggests that development of a smart device or smartphone application that can detect guide dog waste would greatly benefit guide dog users.

\begin{figure}
    \centering
     \includegraphics[width=0.475\textwidth]{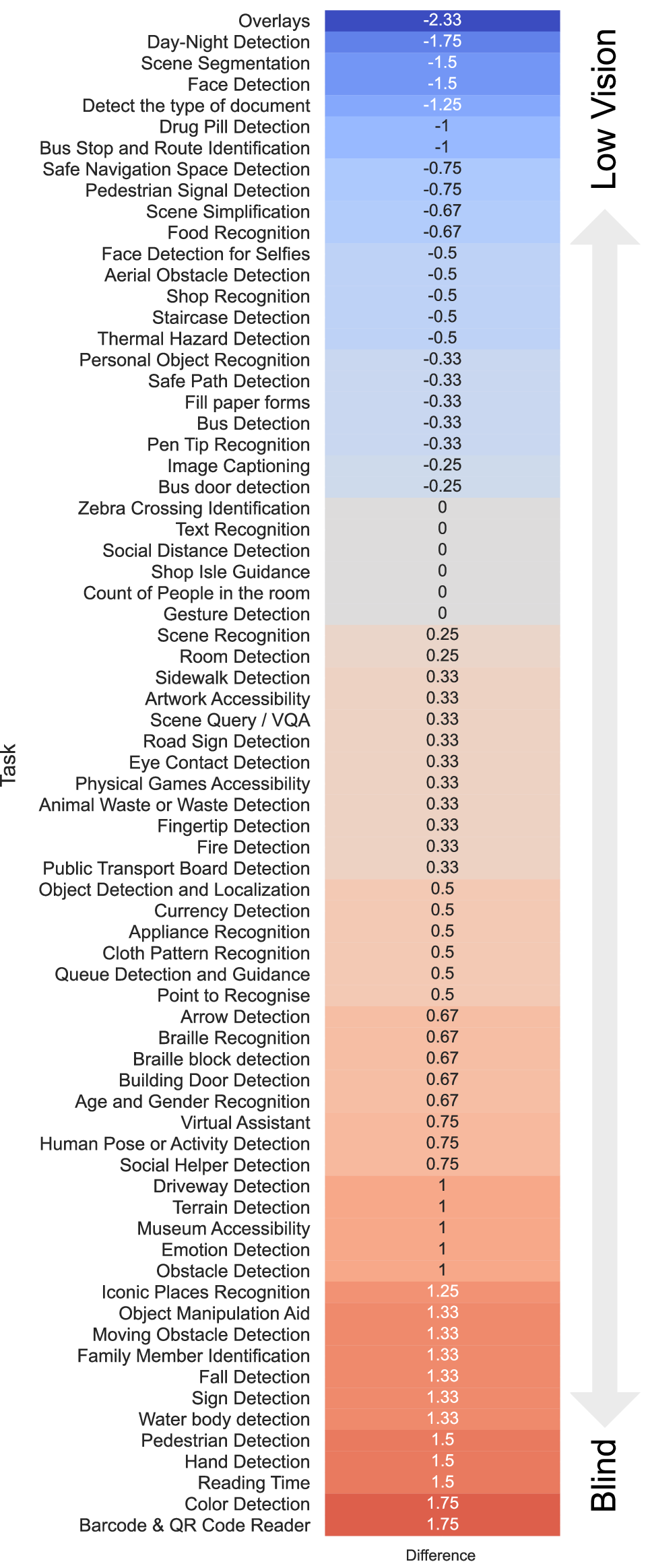}
     \caption{Mean difference of usefulness score of tasks between the blind and low-vision groups.}
     \Description{This figure shows a single column heat map. The colour transition from blue at the top and red at the bottom. The blue at the top indicates the tasks preferred more by the low-vision participants and the red at the bottom indicated the tasks preferred more by the blind participants. The trends of the plots are described in the fourth paragraph of the Ranking of researcher tasks subsection in Section 4.3 Results. Refer to Figure8.csv in supplementary material for all the data.}
     \label{fig:percentage-difference-usefullness}
\end{figure}

We conducted an analysis to measure if low-vision and blind participants had different task rankings.
Figure \ref{fig:percentage-difference-usefullness} presents the mean difference in scores assigned by the two groups.
Our findings indicated that low-vision participants assigned greater importance to Overlays and Scene-Segmentation, whereas blind participants favoured Barcode \& QR Code Reader and Color Detection.
The explanation of Overlays and Scene-Segmentation is straightforward for low-vision participants as they can depend on the visual mode of interactions.
Blind participants mentioned that a device capable of identifying and reading Barcode and QR codes would be more useful as they are not able to locate them.
All four blind participants mentioned that Color Detection is invaluable (3) or very useful (1) especially if it was accurate even under different lighting conditions, compared to the two low-vision participants mentioning it will only be slightly useful.
Additionally, our analysis identified that Text Recognition tasks exhibited no significant difference between the two user groups.
This, in conjunction with the results from the previous section regarding the participants preferred tasks, highlights the necessity for accurate text recognition devices regard of the visual condition.
However, it is important to note that the sample size for this comparison  was extremely limited, with only six participants ranking each task and only 2-3 of them being low-vision participants.

Based on the `best' type of device selected by the participants for each task, we calculated the count of all the wearables as shown in Column C in Figure \ref{fig:wearable-types}.
One thing to note is, participants were allowed to select more than one `best' wearable for each task.
For example, one participant gave hand-held device and smartphone for Color detection.
Based on this, head-mounted devices, smartphones, and hand-held devices were the top three device types for each of the tasks.
We also found no discernible preference in the type of device between the blind and low-vision participants.

Finally, six out of the 24 participants chose to update their top five list when given the opportunity to do so after ranking the tasks.

\subsubsection*{\textbf{Overall perception of smart devices:}}
All participants expressed that they considered computer vision based smart devices to be highly valuable for their future needs. 
P22 summarised this clearly:
\begin{quote}
    \textit{"Because of the independence they offer a person who is blind or visually impaired to act as normally as possible. In external environment, it build somebody's confidence and sense of well being. In a home environment, it would mean not relying on others all the time to assist with visual tasks and it would increase safety and save alot of time."}
\end{quote}

Subsequently, we inquired whether participants had any concerns or design considerations that the research community should take into account while creating AI-enabled smart devices.
Of the 24 participants, 16 cited ease of use as their top concern, encompassing factors such as longer battery life, reliability, and portability.
Another 10 participants emphasised the need for smart devices to be subtle and visually appealing. 
As P5 encapsulated this sentiment:
\begin{quote}
\textit{Nobody wants to walk around in Big Ski Goggles, because you feel conspicuous, and everybody's staring at you for a different reason, not just for being blind, because you're in these great big, ugly glasses.}
\end{quote}
The majority of these current limitations of smart devices are due to hardware constraints, which were discussed in previous studies \cite{golubova_design_2021}.
However, it is reasonable to anticipate that these limitations will be addressed as the technology evolves, resulting in devices that are both smaller in size and more powerful.
\section{Comparison \& Discussion}
\label{section: analysis}

In this section we contrast the results of the two studies and also discuss the implications for smart device research and adoption.

\subsection{\textbf{Which tasks should researchers explore?}}

We were interested to see the alignment between researcher focus on tasks as revealed by the number of papers addressing the task in Study 1 and the BLV participants' ranking of the importance of the task found in Study 2. We first computed the correlation between the two. This was \emph{r}(70) = .14, \emph{p}=.24. indicating a weak positive correlation.

\begin{figure*}
    \centering
    \includegraphics[width=\textwidth,keepaspectratio]{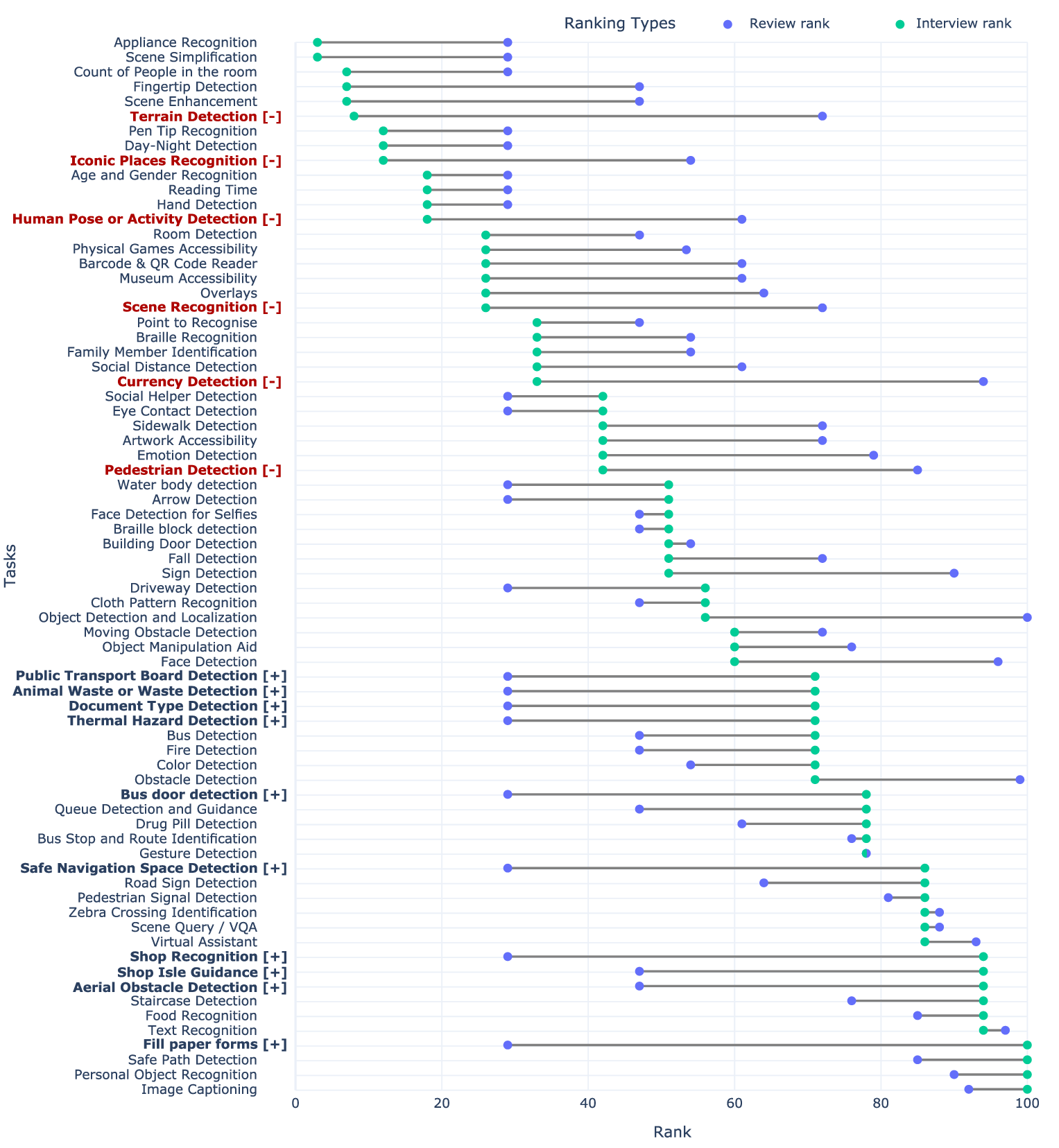}
    \caption{Ranking of various tasks based on the number of papers in the review [Review rank] and interview preference [Interview rank]. Tasks that were \textbf{important} to the participants, but were \textbf{not the focus} of the researchers are highlighted in blue with a [+] symbol next to them and tasks that were deemed \textbf{unimportant} by the participants, but were the \textbf{focus} of the researchers are highlighted in red with a [-] symbol next to them.}
    \Description{This figure shows a graph that shows the distance between the review rank and interview rank for each task. The two ranks are plotted on a horizontal axis and connected with a line. The tasks are sorted by the lowest to highest interview rank. Refer to Figure9.csv in supplementary material for all the data.}
    \label{fig:task-ranking}
\end{figure*}

We then investigated the differences for each task. The tasks identified in Study 1 were ranked based on the count of studies identified for each task (named \textit{"Review Rank"}). 
In Study 2, tasks were ranked based on the perceived usefulness of the tasks by the participants (named \textit{"Interview Rank"}). We scaled and plotted these two rankings in Figure \ref{fig:task-ranking} to explore differences in the ranking. This reveals:

\begin{itemize}
    \item Tasks that were \textbf{important} to the participants, but were \textbf{not the focus} of the researchers (highlighted in Blue with a [+] symbol)
    \item Tasks that were deemed \textbf{unimportant} by the participants, but were the \textbf{focus} of the researchers (highlighted in Red with a [-] symbol)
\end{itemize}

\subsubsection*{Tasks that were \textbf{important} to the participants, but were \textbf{not the focus} of the researchers:} \label{sec:participant_imporant_in_discussion}

Although filling paper forms had the lowest number of studies, we found that all six participants who ranked it, gave a higher usefulness score (Invaluable (5 participants) and Very Useful (1 participant)). 
In fact, only one group of researchers have worked on the filling paper forms task, as reported in \cite{paperform0, paperforms1, paperforms2}. 
Interestingly, these authors found that all of their participants stated they would use such a device in their everyday life.
Considering our findings, we suggest researchers should be prioritising tasks that have been highlighted in blue, such as Aerial Obstacle Detection, Shop Recognition, and Safe Navigation Space Detection.

Although we have highlighted few specific tasks based on the existing research, we acknowledge that other tasks may also be significant, as indicated by the bolded text in Figure \ref{fig:user-preference}. 
Participants in our study identified tasks such as Empty Seat Detection, Sorting Items, and Checking Appearance as important. 
Additionally, researchers can also explore tasks for which BLV users currently rely on sighted assistance, as described in Section \ref{sec:current-technology-use}. 
Therefore, we suggest that researchers consider both existing research and participant feedback when identifying significant tasks for smart devices aimed at aiding the BLV community.

\subsubsection*{Tasks that were deemed \textbf{unimportant} by the participants, but were the \textbf{focus} of the researchers:}
Tasks such as Terrain Detection, Currency Detection and Pedestrian Detection were among the most extensively studied in the literature.
However, participants reported that they did not consider these tasks to be important. 
For instance for Terrain Detection, P7 stated that they could identify terrain by the sensation under their feet, while P10 noted that others would warn them about slippery conditions.
For Pedestrian Detection, participants who used guide dogs indicated that the dogs are good at avoiding collisions, while others reported that they could sense when someone was approaching through hearing or tactile cues. 
Participant P24 offered additional insight into this:
\begin{quote}
    \textit{"I think as a cane user that would be useful, [...] because I'm a dog user, it would actually be not so useful because the the dog detect those people and walks around them, or people see you coming, and they might move out of the way."}
\end{quote}
It is also important that these devices needs to gain the trust of the BLV community as described by \cite{plikynas2020indoor}.
So smart devices should be designed to work in tandem with the currently established orientation and mobility standards, particularly as these technologies continue to evolve and mature.

Our analysis also revealed that Currency Detection exhibited the greatest disparity between the literature and the perceptions of our participants. 
However, it is important to note that the majority of our participants were from Australia, a country where (in particular since the pandemic) cashless transactions are the norm, rendering the ability to identify individual notes and coins less critical. Furthermore Australian currency has been designed to be tactually distinct.
To gain a more comprehensive understanding of the importance of Currency Detection among individuals who rely heavily on cash transactions, future studies should consider recruiting participants from regions where cash remains a primary form of payment and the currency may be difficult to distinguish tactually.

\subsection{Which interaction modalities are preferred?}

\subsubsection*{\textbf{Question Answering interaction model as a conversational agent:}}
All 24 of our participants expressed the desire for at least one of their top tasks to use a Question Answering (QA) interaction model. 
This is in line with the increasing popularity of voice input and speech output modalities among BLV participants \cite{voicecontrolled1, speech_input_blind}, and emphasises the importance of smart devices acting as conversational agents.
It also accords with social agency theory \cite{mayer2003social, moreno2001case, converstationalagent}, which suggests that users become more engaged when they are presented with a human-human social interaction.

A conversational interface is particularly suited to more complex tasks such as the second most requested task (Figure \ref{fig:user-preference}), "Scene Query" (9 participants), in which users wish to ask the device questions they might have about their surroundings. 
As P8 expressed, when they want to understand the menu at a restaurant, they prefer to engage in a natural conversation with the device to understand the items on it, as opposed to simply having it read text off the menu.
The ideal device would function as a conversational agent, capable of switching contexts and serving as a personal companion that users can trust. 

Since 2010 \cite{vizwiz1}, the field of question answering for blind people has seen significant exploration. This includes the collection of large datasets of visual questions and answers such as VizWiz \cite{vizwiz2}.
However, our review only identified 13 studies that employed the Question Answering interaction model in smart wearable devices suggesting that this area remains under-explored in this context.

We also note that the emergence of GPT-4 \cite{openai2023gpt4} has led to the introduction of a beta version of Envision glasses, known as "Ask Envision" \cite{envision_2023}, aiming to provide seamless user-device interactions through conversational interfaces. 
This potentially enables question answering functionalities on wearable devices.

Another crucial aspect of the interaction is determining the appropriate amount and timing of information delivery to the user.
A recurring concern among participants was not to be overwhelmed by information and they expressed a desire for devices to only share information that they find interesting/relevant. 
For instance, participant P14 stated, "In an ideal world, you want it [smart device] to read only what I'm interested in."
%In contrast, only 47 studies focused on specific object detection tasks, as shown in Figure \ref{fig:task-counts}.
%We cannot confirm whether this was the intention of researchers, but nonetheless, the ability to determine, display, and provide contextually relevant information to the user is critical.

We also identified that the devices should present information to the user in a manner that is easily understandable, such as using step-based distances instead of meters and feet, and directional indicators such as "to the right" or "at 2 o'clock" instead of cardinal directions like North, East, South, West. 
As participant P1 stated:
\begin{quote}
    \textit{"If a device was giving me direction, that [in form of steps] would be something I would understand easily. [...] If somebody said five meters, I would have to kind of process that, but if somebody said 10 steps that would be more efficient information for me."}
\end{quote}
The majority of studies relied on automated text-to-audio tools to read out the detected item \cite{ashiq2022cnn, yumang2021raspberry, chang2020medglasses}.
Therefore, further exploration is needed into this issue, as only a limited number of studies have focused on how information is conveyed to the user \cite{lu2021personalized, matei2022safety}.

Finally, two participants expressed their preference for the device not to read sensitive information aloud or cause disturbance to those around them. 
This issue has also been identified in previous studies where the majority of participants expressed concerns about privacy in speech-based smart devices \cite{futuremobile, mobiledevice_adoption}.
To address this issue, two potential solutions can be considered: desensitising the device when reading sensitive information aloud or using bone-conducting headphones to ensure privacy.

\subsubsection*{\textbf{How important is the visual modality for low-vision individuals?:}}
Several studies have investigated the potential of image enhancement as an assistive interface for people with low-vision to perceive and comprehend their surroundings \cite{zhao2015foresee, zhao2016cuesee, fox2023using, zhao2019designing}. 
However, our study found that our low-vision participants did not prefer visual feedback, and only 3 (out of the 10 low-vision) participants mentioned they would prefer visual output. 
We found this surprising and it may be because most of our low-vision participants  had only light perception and progressive visual impairments, as discussed in the limitations section.

In a 2020 study by Zhao et al. \cite{zhao2020effectiveness}, they compared the cognitive load between visual and audio feedback during navigation and found that 4 out of the 16 participants reported that the visual feedback was more distracting due to the extra effort required to perceive and understand the visual content, compared to following audio instructions. 
While this finding accords with our findings, we believe that other factors, such as environment (crowded, calm), amount of information, and the specific visual condition, also have an impact on preference.
Therefore, further research is needed to clearly understand the factors that lead to the preference for visual or audio modalities.

\begin{figure}
         \centering
         \includegraphics[width=0.38\textwidth]{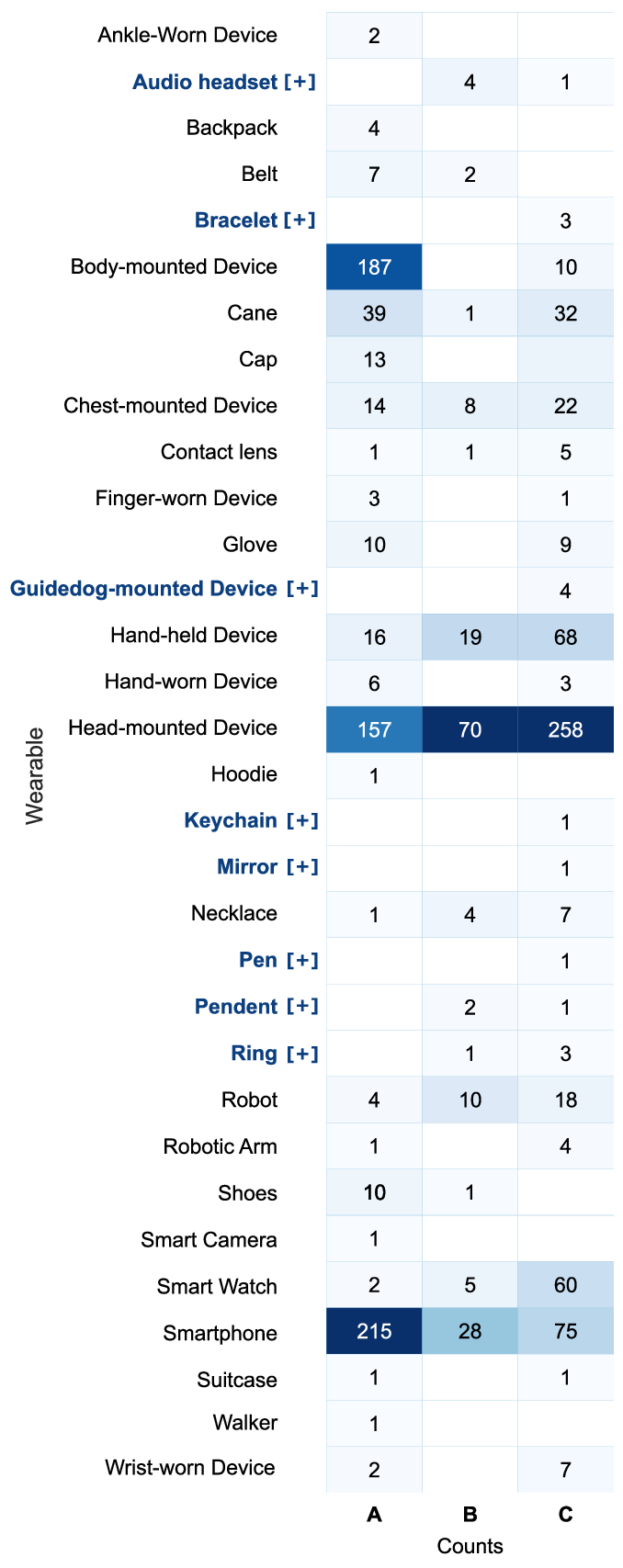}
        \caption{Comparison of Wearable Device Type Counts. (A) Device counts across studies in the review. (B) Count of devices preferred by participants for their top 5 tasks. (C) Counts of devices preferred by participants for different tasks in the review. Devices not found in our review are indicated in bold font with a [+] symbol next to them.}
        \Description{This figure contains a table that is color coded according to the count of the columns, indicating a darker shade for higher counts and lower shade for smaller counts. Refer to Figure10.csv in supplementary material for all the data.}
        \label{fig:wearable-types}
\end{figure}

\subsection{Which devices are preferred?}

Figure~\ref{fig:wearable-types} reveals large differences between the preferences of BLV participants in Study 2 and the devices used by researchers in Study~1. 
We see that researchers have focused on smartphone applications, body-mounted devices and then head-mounted displays. 
However body-mounted devices were rarely mentioned by participants and participants had a strong preference for head-mounted devices over other devices, even if we aggregate hand-held devices and smartphones into a single category.

The authors of \cite{futuremobile} compared the benefits of wearables compared to smartphones and also found that head-mounted devices have the most potential to provide multiple interaction modalities for better accessibility.
This trend aligns with the growing use of head-mounted devices as an assistive tool for the visually impaired, as identified by Li et al. (2022) \cite{li2022scoping}. 

Additionally, we identified wearables that were not included in the review but were mentioned by the participants (highlighted in bold in Figure \ref{fig:wearable-types}). 
One notable wearable type is guide dog mounted wearables, as individuals who rely on guide dogs could potentially use the harness to mount a smart device on it. 
Although no studies were found in the review that utilised such a system, it may be a promising avenue for future research.

We have also observed that the selection of a wearable device depends on the task and the environment in which it is being performed.
P16 mentioned that using a hand-held device or smartphone in public can attract unwanted attention and increase the risk of theft. 
In contrast, a smart glass can blend in as a regular pair of glasses and reduce the risk of snatch and runs (as participants mentioned in this study \cite{futuremobile}). 
However, when at home, a hand-held device or smartphone app would be more appropriate as they are more aware of their surroundings and have both hands free.
Other participants also indicated a preference for hand-held devices for color detection, which is mainly used at home, while smart glasses were preferred for text recognition, which is used in both public and home settings.
Therefore, it appears that the choice of wearable device could be determined by the specific task and the context in which it is used. 

\subsubsection*{\textbf{Universal Device or Platform:}}
The results of our study showed that ease of use is a top concern for blind and low-vision users when it comes to smart devices. 
This is consistent with findings from previous studies \cite{sandnes2016low, lorenzini2020factors, futuremobile}. 
One noteworthy observation was that three participants expressed the opinion that if a particular task can already be performed using an existing device, such as a smartphone, there is no need to develop a separate device solely for that task. 
This sentiment was voiced by participant P4:
\begin{quote}
    \textit{"We already have existing smart devices, so let's reduce the design for all these new devices that are ridiculously priced, people can't afford, and we end up walking around which so much stuff hanging off us."}
\end{quote}

This observation highlights a significant challenge that may face real-world deployment of smart devices. 
Our review revealed that much of the research in this field is conducted in isolated silos, resulting in the creation of separate wearable devices and applications for different tasks. 
Implicitly, this means that users are expected to switch between these devices as needed, which will be inconvenient and they may also need to purchase multiple devices.

We believe there is a pressing need to develop only a few multipurpose devices and applications that seamlessly work together on these devices. 
In our ideal world, a user with visual impairment will be able to utilise a single smart assistive device (probably head-mounted) to gain access to information regarding buses, safely board the bus, locate unoccupied seats, carry out shopping at their intended destination, and participate in social interactions with peers without experiencing any form of stigma.

Finally, an essential aspect highlighted by five participants was the need for adequate training and opportunities to test these devices before making a purchase. 
Given the diverse range of visual conditions that these devices must accommodate, it is advisable to provide potential buyers with the opportunity to test them before making a purchase.
As P2 said:
\begin{quote}
    \textit{"Instead of saying, oh! this [smart-device] is available, go and buy it, they should ask someone that either can come to your home or you can go to the organisation, and have a look at it and [see] if you feel comfortable with it. I'm sure they have things that they can lend out, so that you know to see if it's going to be suited to your needs, because some things aren't suited to your needs, and unless you use them you don't know, because you're not experiencing it. It also makes you feel more confident when you have bought it, because you know it's going to work for you."}
\end{quote}
Tapu et al. \cite{tapu2020wearable} also noted this, highlighting the absence of structured training protocols for individuals with visual impairment when using smart assistive devices.
The authors noted that BLV users often have to rely on self-directed learning without access to a trainer or clear set of instructions.

\subsection{Why is Blind and Low-Vision community involvement important in smart assistive device research?}

As identified by this study, one of the major challenges encountered in current research into smart devices is the mismatch between researcher choice of task and device and the preferences of BLV people. We believe that at least in part this problem arises because of a lack of involvement of BLV people in the research, and, in particular, involvement in the design stage.

To investigate if this is the case, we computed the correlation coefficient between the usefulness scores given by the participants and count of papers categorised from the review as follows:\\

\begin{tabular}[t]{p{5cm}|p{2cm}}
    \textbf{Studies} & \textbf{Correlation Coefficient} \\ \hline
    No BLV participants & \emph{r}(70)=.13,\emph{p}=.28. \\
    With BLV participants only involved in the evaluation stage  &\emph{r}(70)=.16,\emph{p}=.18. \\ 
    With BLV participants involved in the design stage &\emph{r}(70)=.27,\emph{p}=.02. \\ 
\end{tabular} \\[2mm]

This reveals that studies that included BLV participants during the design stage demonstrated a significantly higher positive correlation compared to the other two groups. While correlation does not necessarily mean causation, this does 
suggest that involving BLV participants in the design stage may lead to the development of smart devices that perform useful tasks for the community. 

Nonetheless, it should be noted that overall, there is no strong correlation indicating that BLV participant involvement guarantees the creation of a device that is generally perceived to be useful by the BLV community. We believe identifying the most useful applications of new assistive technologies is best served by studies like this and that of~\cite{golubova_design_2021} that encourage BLV participants to ``brainstorm'' possible applications of these technologies.

A number of other studies have demonstrated the advantages of involving users in the development process, including but not limited to, generating superior ideas, shortening the development cycle, and increasing user satisfaction \cite{steen2011benefits, kujala2003user, empathy}.
Duckett \& Pratt \cite{participant_importance} also found that the BLV community believes that researchers could make a greater contribution to the field by consulting them more thoroughly in research design and practice, and by giving them more control over research activities. 

Our findings suggest that many smart-device studies prioritise development of the technology over the requirements of the user, which could potentially lead to a low adoption rate of these devices and a disconnect between the research community and the intended user base \cite{shivakumar2018digital}. 

We believe it is imperative that researchers recognise the importance of involving BLV participants from the beginning of the research process through exploratory studies and adopting a co-design approach.
However, we note that involving BLV participants  can be challenging due to various factors, such as the difficulty in finding willing participants, ensuring appropriate reimbursement, and addressing ethical considerations~\cite{review_quantitative}. 

Another consideration was raised by \cite{li2022scoping} where they showed that many of the studies on head-mounted-devices were done in a controlled lab environment thereby not reflecting the actual real world use cases.
So more studies involving BLV participants also needs to be done in read world scenarios.
%Nevertheless, given the potential benefits, there is a strong incentive and pressing need for more studies to involve and co-design assistive devices with the BLV community.

\section{Limitations}
\label{section:limitation}

Our study had a limited sample size (24 participants), and only six participants were asked to rank the tasks identified in the review, and only two to three of these had low-vision and only three or four identified as blind. 
This may not be sufficient to draw conclusive results, in particular differences between preferences of participants who are blind and those with low vision, and further studies with larger sample sizes are warranted.
Additionally, the low-vision participants in our study had, on average, visual acuity closer to total vision loss with light perception and had progressive vision loss.
This may have impacted the results, in particular the low preference for visual output. 

%Our analysis of relevant literature indicates that studies on this topic have been conducted in various countries. 
Another limitation is that our study sample was predominantly drawn from a single developed country, with only one participant from a developing country. 
Moreover, the participants were recruited through an electronic mailing list, which may have resulted in a self-selection bias, with individuals who are more technology-savvy being over represented in our sample. 
These factors are reflected in the technology adoption rates of our participants, as presented in Table \ref{tab:participant_details} (in the Appendix).

We acknowledge that our study was limited to research conducted between 2020 and mid 2022. Consequently, if there are significant bodies of works prior to or beyond this period they will not have been included in our analysis.

It is also crucial to acknowledge the rapidly evolving field of smart devices. 
With the introduction of GPT models, such as GPT4 \cite{openai2023gpt4}, devices will soon be capable of capturing visual scenes and answering users' prompted questions. 
This is exemplified in the new features by Envision \cite{envision_2023} and Be My Eyes \cite{bemyeyesvolunteer}. 
Our study did not explicitly consider how BLV users would like to interact with GPT4-enabled devices and further studies are required.
\section{Conclusion}

In this paper, we conducted two studies to evaluate whether current smart assistive technology research utilising AI-based computer vision to understand the immediate environment addresses the needs of blind and low-vision (BLV) users. 
Study 1 involved a scoping review of 646 papers, while Study 2 comprised semi-structured interviews with 24 BLV participants to ascertain their preferences. 

Our findings reveal only a weak positive correlation between the perceived importance of tasks by BLV users and researchers' focus on those tasks. 
Although BLV participant involvement during the research design stage improved the correlation slightly, it remained low. 
Moreover, conversational agent type devices are preferred by BLV participants, but limited research has been conducted on this topic. 
We found that head-mounted devices are preferred by BLV users, although the choice of wearable device depends on the task and environment, and that there was only limited preference for visual output by our low-vision participants. 
Finally, we suggest that researchers focus on more universal device or platform rather than bespoke devices so as to enable seamless usage without the need for device switching. 

We believe this research will provide valuable guidance for researchers working in the field of smart assistive technologies and enable them to focus on tasks and interactions that are prioritised by the BLV community.

\begin{acks}

We are grateful to the Monash Data Futures Institute for providing the funding that made this project possible.
We would like to extend our appreciation to our participants Adua Merola, Cass Embling, Cynthia Kate Gregory, Debra L Simons, Duc Anh Minh Nguyen, Elise Lonsdale, Erin Goedhart, Grace King, Iris Wilson, Janene Sadhu, Jennifer Parry, John Hardie. Josephine Mckinnon, Leonore Scott, Maree Fenech, Marée Steinway, Meredith Prain, Michael Hamon, Neslihan Sari, Peter Mason, Rocco Cutri, Susie Rich, Yvonne Huntley and other participants for their invaluable contribution to our study.
Their active participation significantly enriched the study, and we are truly appreciative of their involvement.

\end{acks}

\bibliographystyle{ACM-Reference-Format}
\bibliography{sample-base}

\appendix

%TC:ignore
\newpage
\section{Tech Adoption Groups}

\begin{table}[htbp]
    \centering
    \caption{Tech Adoption Groups as classified by \cite{beal1957diffusion}}
    \begin{tabular}{|p{3cm}|p{4cm}|}
    \hline
    \textbf{Adoption Group} & \textbf{Description} \\ \hline
    Innovator (I) & A person who builds or develops smart assistive technology \\
    Early Adopter (EA) & A person who buys new smart assistive technology as soon as its available to public \\ 
    Early Majority (EM) & A person who buys new smart assistive technology if its becoming popular with everyone\\ 
    Late Majority (LM) & A person who waits until majority leaves feedback and reviews on the device before buying\\ 
    Laggards (L) & A person who is very conservative and does not like to change their current tools\\
    \hline
    \end{tabular}
    \label{tab:technology-adoption}
\end{table}

\section{Details of participants}

\begin{table}[htbp]
    \centering
    \caption{Details of the participants}
    \label{tab:participant_details}
    \begin{tabular}{|l|l|l|p{1cm}|p{1cm}|p{1cm}|}
        \hline
        & \textbf{Group} & \textbf{Age} & \textbf{Level of Vision} & \textbf{Age of Onset} & \textbf{Tech Adoption} \\ \hline
        P1        & 1        & 55 to 64                     & LV               & Birth                 & LM          \\ \hline
        P2        & 2        & 55 to 64                     & LV               & Birth                 & EM         \\ \hline
        P3        & 2        & 18 to 24                       & LV               & Birth                 & LM          \\ \hline
        P4        & 3        & 45 to 54                     & LV               & 26                    & L               \\ \hline
        P5        & 4        & 55 to 64                     & LV               & Birth                 & EA          \\ \hline
        P6        & 4        & 45 to 54                     & LV               & 17                    & LM          \\ \hline
        P7        & 1        & 25 to 34                     & B                    & 3                     & LM          \\ \hline
        P8        & 1        & 25 to 34                     & B                    & Birth                 & LM          \\ \hline
        P9        & 1        & 65 or over                   & B                    & Birth                 & LM          \\ \hline
        P10       & 1        & 55 to 64                     & B                    & 9                     & EM         \\ \hline
        P11       & 2        & 35 to 44                       & B                    & 16                    & EA          \\ \hline
        P12       & 2        & 45 to 54                       & B                    & 38                    & EM         \\ \hline
        P13       & 2        & 45 to 54                     & B                    & Birth                 & LM          \\ \hline
        P14       & 3        & 35 to 44                     & B                    & Birth                 & EA          \\ \hline
        P15       & 3        & 65 or over                   & B                    & 46                    & LM          \\ \hline
        P16       & 3        & 35 to 44                     & B                    & Birth                 & LM          \\ \hline
        P17       & 3        & 55 to 64                       & B                    & Birth                 & EM         \\ \hline
        P18       & 4        & 45 to 54                     & B                    & Birth                 & EA          \\ \hline
        P19       & 4        & 55 to 64                       & B                    & Birth                 & EM         \\ \hline
        P20       & 4        & 45 to 54                     & B                    & Birth                 & EA          \\ \hline
        P21       & 4        & 65 or over                   & LV               & Birth                 & LM          \\ \hline
        P22       & 3        & 55 to 64                       & LV               & 12-14                 & LM          \\ \hline
        P23       & 2        & 65 or over                   & LV               & 64                    & LM          \\ \hline
        P24       & 1        & 35 to 44                     & LV               & 15                    & EM         \\ \hline
    \end{tabular}
\end{table}

\newpage
\section{Tasks}

\begin{table}[h!]
    \centering 
    \caption{This table gives the tasks extracted from the review process and their grouping into higher-level categories.}
    \label{tab:tasks}
    \begin{tabular}[t]{|p{\linewidth}|}
    \hline
        \textbf{Assistive products for personal mobility} \\
        \hline
        General Navigation\\
        \hline
         Obstacle Detection - \textit{Identify obstacles in the path to navigate safely} \\
         Safe Path Detection - \textit{Detect and recommend safe paths for navigation} \\
         Staircase Detection - \textit{Recognise staircases for navigation} \\
         Moving Obstacle Detection - \textit{Identify dynamic obstacles like gates, revolving doors} \\
         Building Door Detection - \textit{Identify doors on buildings for entry points recognition} \\
         Aerial Obstacle Detection - \textit{Detect obstacles that are head height} \\
         Queue Detection \& Guidance - \textit{Detect queues and provide assistance to wait in line} \\
         Arrow Detection - \textit{Detect and interpret directional arrows for guidance} \\
         Safe Navigation Space Detection - \textit{Identify safe areas for navigation and waiting} \\ 
         \hline
        Indoor Navigation \\
        \hline
         Sign Detection - \textit{Identify signs and signage within indoor environments} \\
         Room Detection - \textit{Recognise different rooms or areas in a building} \\
         Shop Isle Guidance - \textit{Provide navigation assistance within shop aisles} \\
         \hline
        Outdoor Navigation \\
        \hline
         Zebra Crossing Identification - \textit{Identify zebra crossings on roads or streets} \\
         Pedestrian Detection - \textit{Identify pedestrians when walking} \\
         Pedestrian Signal Detection - \textit{Detect pedestrian crossing signal} \\
         Bus Stop and Route Identification - \textit{Recognise bus stops and their routes} \\
         Sidewalk Detection - \textit{Detect and locate sidewalks in urban scenes} \\
         Terrain Detection - \textit{Recognise and categorise different types of terrains} \\
         Road Sign Detection - \textit{Recognise and locate road signs} \\
         Iconic Places Recognition - \textit{Recognise famous landmarks and iconic locations} \\
         Braille block detection - \textit{Identify Braille blocks or access tiles on the ground} \\
         Bus Detection - \textit{Detect approaching or stopped buses} \\
         Animal Waste or Waste Detection - \textit{Locate and identify animal waste or litter} \\
         Bus Door Detection - \textit{Identify bus doors for boarding} \\
         Driveway Detection - \textit{Detect driveways in residential or commercial areas} \\
         Shop Recognition - \textit{Identify and name shops or stores} \\
         Public Transport Board Detection - \textit{Identify and read boards displaying public transport information} \\
         Water Body Detection - \textit{Detect and identify bodies of water such as lakes, rivers, etc.} \\
        \hline
    \end{tabular}
\end{table}

\newpage
\begin{table}[h!]
    \centering
    \begin{tabular}{|p{\linewidth}|}
        \hline
        \textbf{Assistive products for communication and information}\\
        \hline 
        Human Interaction \\
        \hline
         Face Detection - \textit{Locate faces and identify them} \\
         Emotion Detection - \textit{Identify emotions of people} \\
         Gesture Detection - \textit{Recognise hand or body movements pointing, raising hands} \\
         Human Pose or Activity Detection - \textit{Recognise body poses or actions such as sleeping, sitting} \\
         Social Distance Detection - \textit{Measure and help maintain safe/comfortable distances from people} \\
         Family Member Identification - \textit{Locate and Identify family members} \\
         Face Detection for Selfies - \textit{Find faces specifically in selfie-style images to help frame better} \\
         Age and Gender Recognition - \textit{Determine age and gender of people} \\
         Count of people in a room - \textit{Estimate the number of individuals present} \\
         Eye Contact Detection - \textit{Detect whether eye contact is made during conversation} \\
         Hand Detection - \textit{Detect and track human hands to pass and receive objects} \\
         Social Helper Detection - \textit{Identify individuals offering assistance such as policeman} \\
         \hline
         Textual and Graphical Information\\
        \hline
         Text Recognition - \textit{Detect and read text} \\
          Barcode \& QR Code Reader - \textit{Scan and decode barcodes and QR codes} \\
         Braille Recognition - \textit{Convert and read Braille text} \\
         Color Detection - \textit{Identify and detect colors} \\
         Fingertip Detection - \textit{Detect and read text pointed to the finger tip of the user} \\
         Fill paper forms - \textit{Assist with filling paper forms} \\
         Reading Time - \textit{Detecting clock faces to read time} \\
         \hline
        Environmental Information   \\
        \hline
         Virtual Assistant - \textit{Allows user to ask questions about their environment} \\
         Image Captioning - \textit{Generate description of what the user is facing} \\
         Scene Query / VQA - \textit{Allows the user to ask any query and maintain a conversation about their surrounding} \\
         Scene Recognition - \textit{Identify and categorise the type of environment such as a hospital room or a parking lot} \\
         Point to Recognise - \textit{Identify objects or elements pointed at by users} \\
         Scene Enhancement - \textit{Enhances contrast by distinguishing objects with different colours} \\
         Day-Night Detection - \textit{Distinguish whether its day time or night time} \\
         Overlays - \textit{Superimpose digital information or image enhancement onto real-world scenes} \\
         Scene Simplification - \textit{Reduce visual clutter and detail in the user's field of view to enhance comprehension} \\ 
         \hline
    \end{tabular}
\end{table}

\newpage

\begin{table}[h!]
    \centering
    \begin{tabular}[t]{|p{\linewidth}|}
        \hline 
        \textbf{Assistive products for handling objects and devices}\\
        \hline
        General Objects\\
        \hline
        Object Detection and Localization - \textit{Identify and locate various objects} \\
         Object Manipulation Aid - \textit{Assist users in interacting with and manipulating objects} \\
        \hline
        Specific Objects\\
        \hline
         Currency Detection - \textit{Recognise and differentiate currencies for transactions} \\
         Personal Object Recognition - \textit{Identify personal belongings for organisation and tracking} \\
         Food Recognition - \textit{Identify and classify different types of food items} \\
         Cloth Pattern Recognition - \textit{Detect and categorise patterns on clothing items} \\
         Appliance Recognition - \textit{Identify household appliances and their controls} \\
         Pen Tip Recognition - \textit{Detect and locate writing pen tips} \\
         Document Type Detection - \textit{Identify document types such as passports, credit cards, insurance cards} \\
         \hline
    \end{tabular}
\end{table}

\begin{table}[h!]
    \centering
    \begin{tabular}[t]{|p{\linewidth}|}
    \hline
\textbf{Assistive products for personal care and protection}\\
        \hline 
         Fall Detection - \textit{Detect and alert for potential fall incidents in real-time} \\
         Fire Detection - \textit{Detect and alert for fire incidents} \\
         Thermal Hazard Detection - \textit{Identify items with hazardous temperature levels like a hot kettle} \\ 
         \hline
    \end{tabular}
\end{table}

\begin{table}[h!]
    \centering
    \begin{tabular}[t]{|p{\linewidth}|}
    \hline
        \textbf{Assistive products for cultural and sports activities} \\
        \hline 
         Artwork Accessibility - \textit{Detect and provide descriptions of artworks} \\
         Museum Accessibility - \textit{Assist navigation by provide guidance and descriptions of museums artefacts} \\
         Physical Games Accessibility - \textit{Detect and provide descriptions of physical games} \\ 
         \hline
    \end{tabular}
\end{table}

\begin{table}[h!]
    \centering
    \begin{tabular}[t]{|p{\linewidth}|}
    \hline
        \textbf{Assistive products for personal medical treatment} \\
        \hline 
         Drug Pill Detection   - \textit{Recognise and describe medications} \\
         \hline
    \end{tabular}
\end{table}

\end{document}